\documentclass[12pt]{article}             %Arxiv

\usepackage{ifthen} \newboolean{ieee}

\setboolean{ieee}{false} %Arxiv, together with document class above
\ifthenelse{\boolean{ieee}}{}{} %%%%%%%%%% Remove IEEE

\ifthenelse{\boolean{ieee}}{\setlength{\textwidth}{180mm}\setlength{\textheight}{238mm}\setlength{\oddsidemargin}{-8mm}
\setlength{\evensidemargin}{-8mm} \setlength{\headsep}{10mm}
}
{\setlength{\textwidth}{165mm}\setlength{\textheight}{229mm}\setlength{\oddsidemargin}{-1mm}
\setlength{\evensidemargin}{-4mm} \setlength{\topmargin}{-6mm}
}

\makeatletter %%REMOVE FOR SIAP
\renewcommand{\section}{\@startsection
{section} {1} {0mm} {-\baselineskip} {0.5\baselineskip}
{\large\bf}}
\renewcommand{\subsection}{\@startsection
{subsection} {2} {0mm} {-\baselineskip} {0.5\baselineskip}
{\normalsize\bf}} \makeatother

\usepackage{color}

\usepackage{hyperref}

\usepackage{comment} %Para comentar partes 

\excludecomment{versionC}

\usepackage{amsmath}
\usepackage{amssymb}
\usepackage{amsfonts}
\usepackage{amsthm}
\usepackage{latexsym}
\usepackage{epsfig}
\usepackage{array}
\usepackage{boxedminipage}

\newcommand{\vs}{\vspace{1mm}}
\newcommand{\vv}{\vspace{2mm}}

\newcommand{\vvvvv}{\vspace{5mm}}
\newcommand{\vvvvvv}{\vspace{6mm}}

\newcommand{\R}{\mathbb{R}}

\newcommand{\s}{\mathbb{S}}

\newtheorem{defin}{Definition}
\newtheorem{theor}{Theorem}
\newtheorem{lema}{Lemma}
\newtheorem{propo}{Proposition}

\newcommand{\h}{\mathbb{H}}

\newcommand{\ke}{\mathrm{ker} \hspace{0.5mm}}

\newcommand{\tra}{{\sf T}}

\title{%Linear 
Homogeneous Models 
of Nonlinear Circuits\ifthenelse{\boolean{ieee}}{}{\footnote{This is the author's final version
of a paper published in the {\em IEEE Transactions  on Circuits and Systems I - Regular Papers},
vol.\ 67, no.\ 6, June 2020, pp.\ 2002-2015: \url{http://doi.org/10.1109/TCSI.2020.2968306}}
\vs} 
}
\author{Ricardo Riaza\thanks{Depto.\ de Matem\'{a}tica Aplicada a las TIC
\& Information Processing and Telecommunications Center, 
ETS Ingenieros de Telecomunicaci\'{o}n, Universidad Polit\'{e}cnica de Madrid, Spain. {\sl ricardo.riaza@upm.es}.}} 
\date{} %\today}

\begin{document}

\ifthenelse{\boolean{ieee}}{\markboth{IEEE Transactions on Circuits and Systems -- I REGULAR PAPERS
(SUBMITTED)}
%~Vol.~X, No.~Y, Month-Year}%
{Riaza: 
Homogeneous Models of Nonlinear Circuits (SUBMITTED)}}{}

\maketitle

\ifthenelse{\boolean{ieee}}{}{\mbox{}\vspace{-13mm}} %only arxiv

\begin{abstract}
This paper %introduces a new 
develops a general approach to nonlinear 
%electrical and electronic 
circuit modelling aimed at preserving
the intrinsic symmetry of electrical circuits when formulating 
reduced models. The goal is to provide a framework accommodating
such reductions in a global manner and without any loss of 
generality in the working assumptions; specifically, we avoid
global hypotheses imposing the existence of a classical circuit variable 
controlling each device. Classical
(voltage/current but also flux/charge) models are easily obtained
as particular cases of a general homogeneous model.
Our approach extends the results
introduced for linear circuits in a previous paper,
by means of a systematic use of
global parametrizations of smooth planar curves. This makes it possible 
to formulate reduced models
in terms of %so-called {\em 
homogeneous %} 
variables also in the nonlinear context: contrary
to voltages and currents (and also to
fluxes and charges), homogeneous
variables qualify
as state variables %in reduced models of 
for smooth, uncoupled
circuits without any restriction on the characteristics of devices.
The inherent symmetry of this formalism makes it possible to address in broad generality certain analytical
problems in nonlinear circuit theory, such as the state-space problem and 
related issues involving impasse phenomena.
%as well as index analyses of differential-algebraic models. 
%In our setting, the linear/nonlinear distinction in circuit
%theory nicely arises in connection to the different 
%nature of the so-called regular and singular sets in both contexts.
Our framework applies also to circuits with memristors, 
and can be extended to include controlled sources and coupling effects.
Several examples %, %involving Van der Pol oscillators and memristive circuits 
%some of them including memristors,
illustrate the results.
\end{abstract}

%%%%%%%%%[PARA EL ABSTRACT O INTRODUCTION: 
%In topologically nondegenerate problems,
%the homogeneous formalism paves the way for a completely general
%characterization of the so-called regular set in graph-theoretic terms
%(specifically, in terms of the family of spanning trees in the circuit)
%and, subsequently, of the regular manifold
%where the circuit equations define a smooth flow.

%\

%%%%%%%%%%%%[ABSTRACT: 
%nice distinction between linear and (in a strict and local sense) nonlinear circuits
%with regard to the structure of the so-called regular set: empty or the whole space in linear
%cases, open dense in locally nonlinear cases. And the latter are generic]

\ifthenelse{\boolean{ieee}}{{\bf Index Terms:} Analog circuits, circuit analysis, 
nonlinear circuits, memristors, differential-algebraic equations, geometry, network theory, 
nonlinear systems.}{\noindent {\bf Index Terms:} Analog circuits, circuit analysis, 
nonlinear circuits, memristors, differential-algebraic equations, geometry, network theory, 
nonlinear systems.}
%{\bf Keywords:} nonlinear circuit, smooth device, state-space reduction, planar curve,
%closed characteristic, hysteresis, homogeneous coordinates, regular set, impasse set,
%Van der Pol circuit, Murali-Lakshmanan-Chua circuit, memristor.
%nodal analysis, branch-oriented model,
%state model, 
%memcapacitor, meminductor, nodal analysis,
%Manifold of equilibria, normal hyperbolicity, 
%transcritical bifurcation without parameters,  %singularity-induced bifurcation,
%differential-algebraic equation, nonlinear circuit, memristor. 
%qubit, Josephson junction.
%semistate model, 
%matrix pencil, 
%index, tree.
%circuit model, tableau approach,
%modified node analysis, augmented node analysis,
%graph topology, 
%singular ODE, bifurcation, 
%matrix pencil. 
%state-space equation.
%} %only arxiv

%\newpage

\section{Introduction}
\label{sec-intr}

We extend in this paper the %so-called homogeneous 
approach of \cite{homoglin}
to the nonlinear circuit context. Our main goal is to introduce and exploit, for analytical purposes,
circuit models of the form
\begin{subequations} \label{core00}
\begin{eqnarray}
A_c \psi_c'(u_c) u_c' + A_l \psi_l(u_l) + A_{r}\psi_{r}(u_{r}) & = & 0 \\
B_c\zeta_c(u_c) + B_l\zeta_l'(u_l)u_l' + B_{r}\zeta_{r}(u_{r}) & = & 0,
\end{eqnarray}
\end{subequations}
where
we use the prime $'$ to denote differentiation
(with respect to time when no argument is given, as e.g.\ in $u_c'$).
This model is formulated
in terms of certain vector-valued {\em homogeneous variables}, namely 
$u_c,$ $u_l$
and $u_r$ for (smooth, possibly nonlinear) capacitors, inductors and resistors.
Independent sources can be handled jointly with resistors
%in the right-hand side, but also, alternatively, 
%within 
just by rewriting the maps $\psi_r$ and $\zeta_r$  as $\psi_r(u_r, t)$ and $\zeta_r(u_r, t)$,
respectively; memristors can also be easily included in the model and will
be considered later. 

To get a brief overview before going into details, the reader may think
of the matrices $A=(A_c \ A_l \ A_r)$ and $B=(B_c \ B_l \ B_r)$
as describing the circuit topology (with Kirchhoff laws reading as $Ai=0$, 
$Bv=0$), whereas the so-called {\em parametrization maps}
$\psi_c$, $\zeta_c$, etc.\ %$\psi_x$ and $\zeta_x$, 
comprise the characteristics of the circuit devices. 
%as well as 
%$\psi_l$, $\psi_r$ 
%and 
%the homogeneous variables
%$u_x$, 
%$u_c$, $u_l$, $u_r$ 
%are now vector-valued. 
Solutions in terms of classical circuit variables (current, voltage, charge and flux)
are explicitly obtained from those of (\ref{core00}) by means
of the relations $i_r=\psi_r(u_r)$, $v_r = \zeta_r(u_r)$, $\sigma_c=\psi_c(u_c)$, etc. 
%Details in this regard
%are carried out in 
(cf.\ subsection \ref{subsec-main}). %: cf.\ 
%the relations (\ref{homog-res2}), (\ref{homog-cap2}), (\ref{homog-ind2}) and the derivation of the model
%(\ref{core}). 
%In particular, m
%It  is also worth emphasizing 
Note also that models formulated in terms of %some of these 
classical circuit variables 
%(voltage, current, flux and charge) 
are %included in 
%covered by 
%captured in
comprised in 
(\ref{core00}) as particular cases, being obtained 
in a straightforward manner by means of specific choices of the 
%parametrization 
maps $\psi_c$, $\zeta_c$, etc.: the idea is that, 
for example, a voltage-control assumption for all resistors is captured in the model
by setting $\zeta_r=\mathrm{id}$, so that $u_r$ amounts to the vector of branch voltages $v_r$
and then
$\psi_r$ 
describes the voltage-to-current characteristics. We emphasize that the
scope of (\ref{core00}) extends beyond these particular 
(classical) cases, %. This way, (\ref{core00}) 
providing a truly general
and flexible framework for nonlinear circuit modelling and analysis.
We refer the reader to Section \ref{sec-homogmodels}
for further details.

A key element in our approach is the parametric form of Ohm's law, 
which reads as
%writing  Ohm's law in parametric form, that is, %\hspace{-3mm}
\begin{subequations} \label{par-ohm}
\begin{eqnarray}  
i & = & pu \\
v & = & qu.
\end{eqnarray}
\end{subequations}
Here we are dealing with an individual device (a linear resistor) so that
all variables and parameters in (\ref{par-ohm}) are scalar.
We deliberately avoid the current-controlled form $v=zi$ 
($z$ is either the impedance or the resistance, depending on the context)
and the voltage-controlled one
$i=yv$, because both lack generality: indeed,
the former does not accommodate an open-circuit (governed by the relation 
$i=0$),
and the latter excludes a short-circuit (for which $v=0$).
However, all cases are covered in terms of the parameters
$p$ and $q$ in (\ref{par-ohm}),
which are assumed not to vanish simultaneously and therefore
define homogeneous coordinates of a projective line (cf.\ \cite{homoglin});
under the obvious non-vanishing
assumptions, %on $p$ or $q$, 
we get either the impedance/resistance
%in AC sinusoidal steady state) 
%of the corresponding
%linear device 
%in that branch
in the form
$z=q/p$ or the admittance/conductance as $y=p/q$.
%This means that, contrary to (\ref{par-ohm}), 
%classical descriptions are not well-suited e.g.\
% for general parametric analyses
%in which the resistance is intended to be treated as a 
%parameter but the extremal cases
%of a short- and an open-circuit need be included.
%This is relevant in symbolic analysis of circuits, which avoids fixing specific values for the
%circuit parameters.
%Needless to say, %this is relevant in practice because 
%this is the core of an
%idea which arises, in more complex forms, 
%in real problems involving nonlinear and coupled devices, reactive elements, 
%etc.
%Note, additionally, that %i
In (\ref{par-ohm}), 
$u$ is an abstract (so-called {\em homogeneous}) 
variable which will qualify
as a state variable in all possible parameter scenarios,
by contrast to both $i$ and $v$, 
%when used as controlling variables, 
in light of the excluded
configurations resulting from the aforementioned classical forms of Ohm's law.

%(including the extremal cases 
%of a short-circuit, for which $q=z=0$, and an open-circuit, defined by
%$p=y=0$). 
%Sources can be accommodated in a natural manner. 

%Our main goal in this paper is to extend the homogeneous framework
%supporting %models of the form 
%(\ref{linearmodel}) 
%in order to accommodate
%nonlinear circuits.
%Driving the approach to the nonlinear context also 
The extension of this idea to the nonlinear context proceeds
through
%This will naturally involve %requires considering 
the nonlinear counterpart of (\ref{par-ohm}); that is, 
we would now describe %describing
the characteristic of a nonlinear resistor as
\ifthenelse{\boolean{ieee}}{}{\vspace{-6mm}}
\begin{subequations}  \label{nonl-ohm}
  \begin{eqnarray}
    i & = & \psi(u) \\
    v & = & \zeta(u),
  \end{eqnarray}
\end{subequations}
for certain nonlinear functions $\psi$, $\zeta$ and a given parameter
$u$. 
%In Section \ref{sec-param}, we will use 
The key fact here is that this description is %globally 
feasible in a global sense 
for (smooth and uncoupled) nonlinear devices, as a result of
the classification theorem for smooth planar curves. 
%(details are given later) 
This way we will describe the characteristic of each individual device, under 
a smoothness assumption to be made precise later,
in terms of a 
globally defined parameter $u$, lying either on the real
line $\R$ or on the 1-sphere (circle) $\s^1$; this parameter brings
to the nonlinear context the idea of a homogeneous variable discussed above.
Here we 
are assuming that the device is a resistor (in other words,
that its characteristic relates current and voltage), but
the same applies in a natural manner to reactive devices, whose characteristics involve
either
the electrical charge or the magnetic flux, and also to memristors.
%the extension to reactive devices is straightforward.

These ideas are presented in Section \ref{sec-homogmodels} where,
going from the device level of the last two paragraphs to the circuit level,
we derive and discuss in detail the model (\ref{core00}). 
%Specifically, by splitting the circuit elements into resistors,
%capacitors and inductors
%(independent sources can be treated as resistors, and memristors will 
%be introduced later), and using reduced
%cut (or incidence) and cycle matrices $A$, $B$ to capture the
%circuit topology, we will arrive at a model of the form
%\begin{subequations} \label{core00}
%\begin{eqnarray}
%A_c \psi_c'(u_c) u_c' + A_l \psi_l(u_l) + A_{r}\psi_{r}(u_{r}) & = & 0 \\
%B_c\zeta_c(u_c) + B_l\zeta_l'(u_l)u_l' + B_{r}\zeta_{r}(u_{r}) & = & 0,
%\end{eqnarray}
%\end{subequations}
%where the maps $\psi_c$, $\zeta_c$, etc.\ %$\psi_x$ and $\zeta_x$, 
%as well as 
%$\psi_l$, $\psi_r$ 
%and 
%the homogeneous variables
%$u_x$, 
%$u_c$, $u_l$, $u_r$ 
%are now vector-valued. We use the prime $'$ to denote differentiation
%(with respect to time when no argument is given, as e.g.\ in $u_c'$). 
In the absence of coupling effects, the vector-valued
maps $\psi_c$, $\zeta_c$, etc.\
%in (\ref{core00}) 
are %well-defined 
guaranteed to exist in a global sense 
by the classification theorem mentioned above,
having a (say) diagonal form
(that is, the $k$-th component of each map
depends only on the $k$-th component
of its argument); note also that coupling effects may be naturally
accommodated in (\ref{core00}) by deflating this diagonal requirement (cf.\ subsection \ref{subsec-coupling}).
 %in this regard).
%Regarding the model (\ref{core00}), note also that i
Independent sources can be included just by letting
%handled jointly with resistors
%just by rewriting the maps 
$\psi_r$ and $\zeta_r$  depend also on $t$ (e.g.\ for an ideal independent source
injecting a current $i_s(t)$
just set $\psi(t)=i_s(t)$ and $\zeta(u)=u$);
%as $\psi_r(u_r, t)$ and $\zeta_r(u_r, t)$, respectively;
%(provided that they depend also on $t$ in cases beyond the DC setting),
the extension is straightforward and we exclude independent sources throughout the document only for the
sake of brevity.
%something that we assume throughout the document, in most cases
%without %further 
%explicit mention. 
Dependent sources can be handled in a similar manner to coupled devices,
as discussed 
%indicated 
in subsection \ref{subsec-coupling}.

%As detailed 
Even if a detailed discussion can be found in Section \ref{sec-homogmodels}, 
we summarize here some advantages of our homogeneous approach.
Models of the form (\ref{core00}) make it possible to get rid 
of unnecessarily restrictive assumptions
%are of interest from two different perspectives. On the one hand, they 
on controlling variables for the different circuit devices,
much as in the linear case the homogeneous formalism avoids the need to impose 
an impedance (current-controlled)
or an admittance (voltage-controlled) description for each individual device.
And even if
explicit descriptions in terms of a classical circuit variable (current, voltage, charge or flux)
%need eventually be NO THIS IS NOT EXACTLY THE IDEA
are used, all such descriptions can be accommodated in 
(\ref{core00}), which this way defines a broadly general theoretical framework 
for nonlinear circuit modelling.
%avoiding unnecessarily restrictive working assumptions.
Note that the homogeneous setting also allows one to handle, in a global manner, reduced
models 
%(*involving one state variable per branch *REMOVE HERE, EXPAND SOMEWHERE: REDUCTION*) 
in situations in which certain
devices do not admit %such 
a global
explicit description in terms of a classical circuit variable:
an example of this, involving a hysteresis loop, can be found in subsection
\ref{subsec-vdp}. 
In Section \ref{sec-state} %and \ref{sec-bif} 
we apply the homogeneous framework %sketched above
to address certain analytical problems in nonlinear circuit theory, 
involving the state-space problem and also the structure of the so-called
regular and impasse sets. 
\begin{versionC}
** REMOVE, DRIVE SOMEWHERE: For simplicity we restrict the analysis to topologically nondegenerate problems,
the homogeneous formalism paving the way for a completely general
characterization of the so-called regular set in graph-theoretic terms
(specifically, in terms of the family of spanning trees in the circuit)
and, subsequently, of the regular manifold
where the circuit equations define a smooth flow.
Our framework also yields a
nice distinction between linear and (in a strict and local sense) nonlinear circuits
with regard to the structure of the regular and impasse sets. ** UP TO HERE 
\end{versionC}
%: empty or the whole space in linear
%cases, open dense in locally nonlinear cases. And the latter are generic]
We extend in less detail the
results to the memristive context in Section \ref{sec-mem}. 
%Several examples are discussed through all these sections. 
Finally, concluding remarks can be found in Section \ref{sec-con}. 

\section{Homogeneous modelling}
\label{sec-homogmodels}

\subsection{Linear circuits}

The homogeneous formalism in the linear setting is developed in \cite{homoglin}, 
and naturally drives parametric analyses of linear circuits to the context of projective
geometry (related ideas can be found in
%the reader is referred to 
\cite{OtroBryantProjective,
chaiken, penin, smith72}, although none of these works extend
the results to the nonlinear context). 
%for other approaches close to this perspective.
This framework leads to %paves the way for
a completely general reduction of linear circuits, without any restriction on the controlling
variables of individual devices, and to a compact way of writing the equations of any uncoupled
circuit.
% in terms of a vector of homogeneous variables.
In the linear setting, this reduction has 
the form
%yields models with 
%the 
%a truly
%compact form, namely
%\begin{subequations}
\begin{eqnarray} \label{linearmodel}
\begin{pmatrix} AP \\ BQ \end{pmatrix} u =
\begin{pmatrix} AQ \\ -BP \end{pmatrix} \bar{s},
\end{eqnarray}
%\end{subequations}
where $u$ is a vector of homogeneous variables, one for each circuit branch;
$A$ and $B$ are digraph matrices describing the circuit
topology, $P$ and $Q$ comprise
the %aforementioned homogeneous 
parameters $p$, $q$ (cf.\ (\ref{par-ohm})) of individual devices,
either in the real or in the complex setting,
and finally
$\bar{s}$ captures the contribution of sources.
Find details in \cite{homoglin}, where
different analytical properties
of linear circuits are examined %in \cite{homoglin} 
from this perspective. Worth emphasizing if the fact that
classical reductions (not only the branch-voltage and branch-current
models \cite{chen} but also nodal and loop analysis models) can be derived
from (\ref{linearmodel}) by defining regions of the parameter space
which capture different types of working assumptions. For instance, 
a voltage-control assumption, key to the formulation of branch-voltage and nodal models,
is captured in (\ref{linearmodel}) in terms of the nonsingularity of the
$Q$ matrix; in such regions, the model (\ref{linearmodel}) can be naturally recast in 
terms of the voltage vector $v$, or (further) in terms of node potentials. Note 
that it is also possible to combine the homogeneous approach with classical
methods by using a homogeneous formalism only for certain branches, yielding so-called {\em partially
homogeneous} models.

\subsection{Global implicit descriptions of smooth curves. Associate submersions}
\label{subsec-bac}

In the linear context, the formalism above can be understood to rely on the
homogeneous version of Ohm's law, namely
\begin{eqnarray}
  pv-qi=0. \label{homogOhm}
\end{eqnarray}
%with the meaning indicated above for the parameters $p$, $q$ 
Here we are ignoring sources only for %the sake of 
simplicity, %even if 
since they can be easily accommodated 
in the right-hand side of (\ref{homogOhm}). 
As detailed in \cite{homoglin}, a
%circuit device characterized by this equation can be understood to be defined 
resistor governed by (\ref{homogOhm}) can be identified
with a {\em class} of equivalent 
linear forms, namely those which yield the %kernel (or 
zero set % 
in $(i,v)$-space
defined by (\ref{homogOhm}). 
%with the left-hand  side of
%(\ref{homogOhm}) (with specific values
%of $p$, $q$) being only a representative of this class;
%indeed, the kernel of this linear form,
%which is actually the mathematical object
%defining the characteristic curve of the linear
%device, does not change if 
%(\ref{homogOhm}) is multiplied
%by a non-null scalar. 
The key idea is that the $p$, $q$ parameters %in (\ref{homogOhm}) 
are defined only
up to a non-vanishing factor: this naturally frames the linear form in the left-hand side 
of (\ref{homogOhm}), and the resistor itself, in 
a projective line, $(p:q)$ 
%hence 
being homogeneous coordinates of a projective point. 
%This way  the impedance or resistance parameter
%gets %in this formalism 
%a projective meaning. An advantage of this approach
%is that both extremal cases of an open-circuit and
%a short-circuit are included simply by setting $p=0$ or
%$q=0$, respectively. 
%Note that the non-vanishing condition
%on at least one of the derivatives $f_i$, $f_v$ amounts
%in the linear case to requiring that $p$ and $q$ do not
%vanish simultaneously, consistently with the requirement
%that $(p:q)$ define homogeneous coordinates of a projective point.
% Using notions
%from sheaf theory, this approach has been recently extended to the nonlinear context 
%(cf.\ \cite{nonl-subm}).

%\paragraph{Homogeneous incremental resistance.}
%The standard coordinates $(x_1, x_2)$ on $\R^2$ yield two globally
%well-defined linear forms $dx_1$ and $dx_2$ such that 
%$dx_i(v)=v_i$ ($i=1, 2$). %at any $x$
%%: these two forms define the canonical basis of the dual space 
%%$\left(\R^2\right)^{\hspace{-0.3mm}*}$.
%%For later notational convenience, w
%Now, at  
%any given $x \in {\cal C}$ and by means of the 
%the canonical
%identification $T_x \R^2 \sim \R^2$, we choose $(dx_1, -dx_2)$ 
%%(simpler notation than $((dx^1)_x, -(dx^2)_x)$) 
%as a basis for the cotangent space at $x$ (that is,
%the space of linear forms defined on $T_x\R^2$).
%In this basis, the differential at $x \in {\cal C}$ of any submersion $f$ 
%defining ${\cal C}$ % \in S^{\infty}({\cal M},0)$ [OJO ESTA NOTACION]
%has coordinates  
%$\left(f_{x_1}, -f_{x_2}\right).$ 
This idea is extended to the nonlinear context in \cite{nonl-bif}, where a
smooth planar curve defining the characteristic of a nonlinear resistor
is shown to be defined by a family of equivalent 
submersions
(recall that a submersion is a differentiable map  with
an everywhere surjective differential). 
%meaning that the differential $df$ 
%%does not vanish identically anywhere), THIS only for the planar case
%is surjective everywhere). 
The equivalence relation defining these so-called {\em associate} 
submersions, which extends the projective one above, is made precise in \cite{nonl-bif}. 
Given a smooth planar curve,
any such submersion $f$ can be defined on some open subset of $\R^2$ 
including the whole characteristic; it may happen, though, that $f$ cannot be defined on the whole
of $\R^2$. 

Let $f$ be any representative
of the aforementioned equivalence class, that is, consider a smooth planar characteristic
defined by 
\begin{equation}\label{implicit}f(i,v)=0,\end{equation}
for some smooth submersion $f: U \to \R$ defined on an open set $U \subseteq \R^2$.
We may define the {\em homogeneous incremental resistance}
at any point of this characteristic as the pair of homogeneous coordinates
\begin{equation}\label{homincres}
\left(\frac{\partial f}{\partial v}(i, v) : -\frac{\partial f}{\partial i}(i, v)\right),
\end{equation}
whose ratio can be proved independent of the choice of $f$ (find details in \cite{nonl-bif}).
The key aspect of this idea is its global nature: 
%for any smooth planar curve,
$f$ can be defined globally (on some open subset of $\R^2$ including the characteristic) 
%including the whole characteristic; it may happen, though, that $f$ is not defined on the whole
%of $\R^2$. This way 
and the homogeneous incremental resistance so-defined applies
at {\em any} point of the curve, in a way which in essence is independent
of the choice of the submersion $f$ describing the characteristic. %in terms of $f$.
In the linear
case, this definition of the homogeneous resistance
amounts to the aforementioned description as a pair
of homogeneous coordinates $(p:q)$. Note also that we are focusing for simplicity 
on characteristics relating current and voltage but the same applies to those
involving charge and/or flux, so that the same ideas apply to capacitors,
inductors and memristors.

Of course, locally we can always describe a smooth current-voltage 
characteristic
either in terms of the current $i$ or the voltage $v$.
Indeed, 
since $f$ in (\ref{implicit}) is a submersion, 
%(meaning that the differential $df$ 
%%does not vanish identically anywhere), THIS only for the planar case
%is surjective everywhere),
%; in planar cases this is easily checked to amount to
%the requirement that
%the partial derivatives do not vanish simultaneously at any point)
%and in our setting $f$ depends on two variables, it holds that
we know that at every point of the curve at least one
of the partial derivatives in (\ref{homincres}) does not vanish.
%(clearer
%above: smoothness means smooth manifold and
%entails this). 
Fix e.g.\ a point
where the partial derivative $f_v(i, v)$ does not vanish (here we use subscripts for the
partial derivatives for notational simplicity). 
%The implicit
%function theorem supports a 
A local current-controlled description %of the resistor 
%near this point in
%the current-controlled form  
$v=\gamma(i)$ and the expression 
$\gamma'(i)=-(f_v(i, \gamma(i)))^{-1}f_i(i, \gamma(i))$ for the classical incremental resistance follow
naturally from the implicit function theorem.
%We note that the ratio defining
%this resistance does not depend on the choice of $f$. Not so obvious
%Remark also
%that in the linear case this amounts to 
%the classical form of Ohm's law, namely, $v=(q/p)i$,
%where the linear
%resistance is well-defined as the quotient $q/p$
%because of the non-vanishing requirement on $p$.
%In other words, in this setting an admissible choice for
%the homogeneous coordinates defining the homogeneous incremental
%resistance is $(1: \gamma')=(1:-f_v^{-1}f_i)$, where we omit the dependence on $i$
%for notational simplicity.
The same holds
for the %classical 
incremental conductance $\xi'(v)=-(f_i(\xi(v),v))^{-1}f_v(\xi(v),v),$
which is well defined on regions where the partial derivative 
$f_i(i,v)$ does not vanish, allowing
for a local voltage-controlled description $i=\xi(v)$  of the curve.
%(in linear cases, this yields Ohm's law in the form $i=(p/q)v$, with
%the conductance reading as the quotient $p/q$).
But the key remark is that the homogeneous definition (\ref{homincres}), formulated in terms
of the globally-defined submersion $f$, holds at {\em any} point of the
characteristic.

\subsection{Global parametrization of smooth curves and homogeneous
descriptions of nonlinear devices}
\label{subsec-curves}

%\subsection{Smooth characteristics as 1-manifolds in $\R^2$}
%planar curves}
%\label{subsec-bac}

A key question arises at this point, namely, 
how to reduce the implicit description $f(i,v)=0$ (cf.\ (\ref{implicit})) of a smooth
characteristic in terms
of a single variable? Needless to say, this should be relevant in the formulation
of reduced circuit models. We indicated above that this is always feasible in terms
of either $i$ or $v$ {\em in a local sense},
%as a consequence of the implicit function theorem, 
but the goal is to perform such a
reduction in a global manner.
In what follows we show how to do this without the need for additional
assumptions (that is, we will not impose additional conditions supporting
e.g.\ global versions of the implicit function theorem). 
A homogeneous variable $u$ will play the intended global role in the reduction.

As in \cite{nonl-bif}, we assume that the characteristics of the different
circuit devices will be defined by smooth (meaning, for simplicity,
$C^{\infty}$), 
connected 1-manifolds in
$\R^2$. More precisely, they will be {\em regular submanifolds} of $\R^2$,
that is, we assume that 
around every point of the curve there exists a 
so-called adapted coordinate chart relative to the curve: 
the key geometric idea behind this notion is
that the topology induced on the curve
from that of $\R^2$ is such that every point of the characteristic
has a neighborhood
which is diffeomorphic to an open interval;
find details e.g.\ in \cite{tu}.
In this context, the key result making it possible
%As indicated in the Introduction, it is possible
to extend to the nonlinear context the homogeneous description 
(\ref{par-ohm}) 
presented above for linear devices
is %by making use of 
the classification
theorem for smooth 1-manifolds (see
\cite[Appendix]{milnor}). % or \cite[Probl.\ 15.13]{lee}).
This theorem says that any smooth, connected
1-manifold (without boundary) is diffeomorphic either to the real line $\R$ or to the 1-sphere $\mathbb{S}^1$. 
%(we use the latter notation as an abbreviation for $\mathbb{S}^1$). 
This means that
%in our working setting, any characteristic can be described by means of a global smooth parametrization of the form
any smooth planar curve (throughout the document we will 
assume all curves to be connected, without further explicit mention)
can be globally parametrized in the 
form $x=\Gamma(u)$, with $u$ taking values either on the real line $\R$ or on the 1-sphere $\mathbb{S}^1$.
The parameter $u$ 
will play the role of a homogeneous variable in the nonlinear context.
% and we refer the reader to subsection \ref{subsec-homogvar} below for further remarks in this regard.

Later on we will write $\Gamma(u)$ as $(\psi(u), \zeta(u))$ where, for any $u$, 
either $\psi'(u)$ or $\zeta'(u)$
(or both) is (are) non-zero. Note also that, above, we are letting $x$ denote generically a point in $\R^2$:
for different types of devices $x$ will stand either for $(i,v)$ (for resistors) or for
other pairs of variables involving the charge $\sigma$ and/or the flux $\varphi$ (for reactive
devices and, eventually, memristors), as detailed in what
follows.
 
%Here $h$ is a smooth map with values on $\R^2$ and
%satisfying $h'(u) \neq 0$ for all $u$.
%Find additional remarks in this direction in subsection
% \ref{subsec-homogvars}.

\ifthenelse{\boolean{ieee}}{{\it Resistors.}}{\paragraph{Resistors.}} %PTE RESERVAR LOS SUBINDICES PARA PARTE VECTORIAL 
Let us first focus 
the attention on a resistor
%a smooth planar curve ${\cal C}$
%which is assumed to stand for the characteristic of a 
defined by a smooth planar characteristic. 
The classification theorem for 1-manifolds 
implies that there exists
a global parametrization of this characteristic curve of the form
\begin{subequations} \label{homog-res}
  \begin{eqnarray}
    i & = & \psi(u) \\
    v & = & \zeta(u)
  \end{eqnarray}
\end{subequations}
with $\psi'(u)$, $\zeta'(u)$ not vanishing 
simultaneously for any value of the homogeneous variable
$u$. As indicated above, this variable takes values 
either on $\R$ or on $\mathbb{S}^1$. 
%and that $\zeta$ and $\psi$ stand for the components
%of the map $h$ mentioned above. 

%\paragraph{Linearization.} 
The homogeneous incremental resistance 
(\ref{homincres})
%as defined in subsection \ref{subsec-bac}, 
can be naturally
recast in terms of the %homogeneous 
description (\ref{homog-res}),
as shown below.
%In subsection \ref{subsec-nonlindev} 
%We begin by recasting 
%the homogeneous incremental resistance 
%in terms of $\zeta$, $\psi$ and the homogeneous variable $u$;
%recall that this notion was introduced in Definition
%\ref{defin-homogresist} in terms of an arbitrary submersion $f$ 
%describing implicitly the curve.

\vs

\begin{propo}
The homogeneous incremental resistance of a smooth resistor
at a given point $(i,v)=(\psi(u), \zeta(u))$
of the characteristic can be written as
\begin{equation} \label{homresbis}
(\psi'(u): \hspace{1mm} \zeta'(u)).
\end{equation}
\end{propo}

\vs

\noindent
Indeed, let $f(i,v)=0$ stand for the characteristic of the %smooth 
resistor.
By writing $f(\psi(u), \zeta(u))=0$ we get, by the chain rule and using subscripts to denote
partial differentiation,
$$f_i(\psi(u), \zeta(u))\psi'(u)+ f_v(\psi(u), \zeta(u))\zeta'(u) =0,$$ 
so that
$(\psi'(u):\zeta'(u))=(f_v(\psi(u), \zeta(u)): -f_i(\psi(u), \zeta(u))),$
meaning that the ratios are the same; in other words,
the pairs (\ref{homincres}) and (\ref{homresbis})
of homogeneous coordinates describe the same projective point, as claimed. 
%and the claim 
%then 
%follows.
%from (\ref{homincres}).

%Indeed, note that 
%for any $u$ the vector $(\psi'(u), \hspace{1mm} \zeta'(u))$ 
%spans the tangent space
%to the characteristic curve ${\cal C}$ 
%at $(\psi(u), \zeta(u))$. 

%For notational convenience, w
We introduce in the nonlinear context the incremental parameters $p$, $q$ 
%for a resistor 
as
\begin{eqnarray} \label{pqu}
p(u)=\psi'(u), \ q(u)= \zeta'(u),
\end{eqnarray}
so that the homogeneous incremental resistance reads, at any
point of the characteristic, as %\begin{equation}
$(p(u): q(u)).$
%\end{equation}
In the linear
context these amount to the linear coefficients $p$, $q$
arising in (\ref{par-ohm}).
%aforementioned matrices $P$, $Q$.
In these terms, the (classical) incremental
resistance and the  incremental conductance at a given
$u$ read as $q(u)/p(u)$ and $p(u)/q(u)$
(under a nonvanishing assumption on $p$ or $q$, respectively). 
%Mind the notational abuse $p(u)$ vs.\ $p(v, i)$ in
%subsection NO; I AVOID USING P(V, I)
%We note also that the strict local passivity can be also defined in terms of the
%homogeneous variable $u$ simply by requiring that
%both $p(u)$ and $q(u)$ are non-null and of the same sign. 
%;
%in this setting, both the classical incremental
%resistance $q(u)/p(u)$ and the 
%incremental conductance $p(u)/q(u)$ are well-defined and positive.

\ifthenelse{\boolean{ieee}}{{\it Reactive devices.}}{\paragraph{Reactive devices.}} %Nonlinear
Capacitors and inductors defined by smooth characteristics
also admit descriptions in terms of homogeneous variables.
A capacitor with a smooth charge-voltage characteristic 
admits, in light of the aforementioned classification theorem, 
a global parametrization of the
form
\begin{subequations} \label{homog-cap}
\begin{eqnarray}
\sigma & = & \psi(u) \\ v & = & \zeta(u).
\end{eqnarray}
\end{subequations}
We will set $p(u)=\psi'(u)$, $q(u)=\zeta'(u)$ also
for capacitors. We note in passing that
 $p$ and $q$ stand for the derivatives $\psi'$ and $\zeta'$
(cf.\ (\ref{pqu}))
for all types of devices, but that a difference is made by the fact that e.g.\ $\psi(u)$ defines
the current in the resistive case described in (\ref{homog-res}) but the charge in the capacitive
setting (cf.\ (\ref{homog-cap})).
Near points where $q(u) \neq 0$, the capacitor can be locally
described in a voltage-controlled form, with incremental 
capacitance $p(u)/q(u)$. Dually, a charge-controlled description is locally
feasible if $p(u) \neq 0$.
%[Maybe ``Inverse capacitance'' (careful name, does not presume capacitance well-def) is $q/p$ if $p \neq 0$ (and charge-controlled description well def in this case).]

Analogously, for smooth inductors there exists a global
parametrization of the form
\begin{subequations} \label{homog-ind}
\begin{eqnarray} 
i & = & \psi(u) \\  \varphi & = & \zeta(u).
\end{eqnarray}
\end{subequations}
Again, by
setting $p(u)=\psi'(u)$, $q(u)=\zeta'(u)$
we get the incremental inductance in the form
$q(u)/p(u)$ near points of the characteristic where $p(u) \neq 0$,
allowing for a local current-controlled description of the device; as before, 
local flux-controlled descriptions exist near points where $q(u) \neq 0$.
%[Maybe 
%``Inverse inductance'' (careful name, does not presume inductance well-def) is $p/q$ if $q \neq 0$ (and flux-controlled description well def in this case).]

\ifthenelse{\boolean{ieee}}{{\it Classical descriptions.}}{\paragraph{Classical descriptions.}} As indicated in the Introduction,
in addition to accommodating
devices which do not admit a global description in terms of a classical circuit
variable (current, voltage, charge or flux; find an example in subsection \ref{subsec-vdp}), 
the formalism above can also be useful in classical contexts, specifically
when one does not wish to specify in advance %which one is 
the controlling variables for the different devices (e.g.\ for theoretical
purposes or symbolic analysis), 
even if classical descriptions %will
are to be used
eventually. %be 
%used.
%assumed to exist. 
For instance, for nonlinear resistors one can use (\ref{homog-res})
generically, even in the understanding that, when needed, the
description  may
amount to a current-controlled one
(just by setting $\psi \equiv \mathrm{id}$, so that $u$ amounts to the current 
$i$ and $\zeta$ stands for the current-to-voltage function) or to a voltage-controlled
one (by taking $\zeta \equiv \mathrm{id}$, with $u$ standing now
for the voltage $v$ and $\psi$
for the voltage-to-current function). This way the homogeneous formalism
avoids (or delays) unnecessarily restrictive modelling assumptions on 
the characteristics, %of devices, 
making it possible to perform whatever analyses
in broadly general terms.
% and provides a broadly general circuit modelling framework.

\subsection{Homogeneous models of nonlinear circuits}
\label{subsec-main}

%*SOMEWHERE: call $\psi$... PARAMETRIZATION MAPS

%So far our approach is focused on device modelling. 
\ifthenelse{\boolean{ieee}}{{\it Homogeneous description of uncoupled devices.}}{\paragraph{Homogeneous description of uncoupled devices.}}
Extending the framework above from the device level to the circuit level
can be performed in a natural manner under the assumption
that %all devices are smooth and that 
the different group of devices (resistors, capacitors and inductors) do
not exhibit coupling effects (coupled devices are considered
in subsection \ref{subsec-coupling}). As before, we assume that all devices are smooth.

Let us first focus on the description of the resistive devices of a given circuit.
Assume that there are $m_r$ smooth
uncoupled resistors, and let 
$i_r \in \R^{m_r}$
and $v_r \in \R^{m_r}$ stand for the vectors of currents and voltages in the set of
resistive branches.
In the terms detailed in subsection \ref{subsec-bac}, the $k$-th resistor has a 
characteristic which can be written as $f_{r_k}(i_{r_k}, v_{r_k})=0$,
that is, as the zero set of a submersion $f_{r_k}: U_k \to \R$, 
with $U_k$ open in $\R^2$. Altogether, the whole set of resistive characteristics defines a manifold
${\cal C}_r$ of dimension $m_r$ in $\R^{2m_r}$, which is simply the zero set 
$f_r(i_r, v_r)=0$, with the components of 
$f_r$ %having a product structure of $m_r$ 
being the aforementioned individual
submersions 
$f_{r_k}$. Note that the domain of $f_r$ can be written as $U_1 \times \ldots \times U_{m_r}$ %\to \R^{m_r}
after an obvious permutation of variables.
%:\R^2 \to \R$ ($i=1, \ldots, m_r$) %, each one of them
%which describe the characteristic of each resistor, in the terms detailed in Section \ref{sec-equiv}.
Be aware of the fact that the absence of coupling effects confers $f_r$ a simple structure,
since
its $k$-th component depends only on the $k$-th components of the arguments $i_r$ and $v_r$.
Note also that independent 
voltage and current sources can be included in this group of devices in a straightforward manner, 
extending the domains of the corresponding functions $f_{r_k}$ to include time if necessary.

Analogously, the characteristics of the capacitors and inductors define 
two manifolds ${\cal C}_c$ and ${\cal C}_l$, of dimensions
$m_c$ and $m_l$, which can be written
as the zero sets of certain maps %$g_r(v_r, i_r)$,
$f_c(\sigma_c, v_c)$ and $f_l(i_l, \varphi_l)$. %, respectively
In the absence of coupling effects, these maps amount to a product of individual submersions,
as in the resistive case.

%\section{Global parametric description of nonlinear devices}
%{Parametric description of nonlinear devices. Homogeneous variables}
%\label{sec-param}

%\subsection{Linear case}

%The formalism presented in subsection \ref{subsec-lindev} is driven further
%in \cite{homoglin} by means of a systematic use of the parametric form of Ohm's law. Indeed,
%the homogeneous formulation of Ohm's law 
%(\ref{ohmh}) naturally admits an alternative description of each device in parametric form as
%\begin{subequations}
%  \begin{eqnarray}
%    i & = & pu \\
%    v & = & qu.
%  \end{eqnarray}
%\end{subequations}
%Here $u \in \R$ is an abstract variable (called a {\em homogeneous variable}) which allows
%one to eliminate both $i$ and $v$ in the formulation of reduced circuit models, in a way which retains
%the intrinsic symmetry of the theory and without any loss of
%generality (contrary to typical current- or voltage-controlled assumptions...).

%Be aware of the fact that
%the scalar homogeneous variable $u$ is defined only up to a constant, exactly as $p$ and $q$
%are (indeed, the {\em products}
%$pu$ and $qu$ are invariant).

%[MAYBE
%\begin{subequations}
%\begin{eqnarray} \label{res-param}
%    i_{rs}=\psi_{rs}(u_{rs}),    \ v_{rs}= \zeta_{rs}(u_{rs}) 
%\end{eqnarray}
%\end{subequations}
%]

Now, the homogeneous description of individual devices displayed in (\ref{homog-res}),
(\ref{homog-cap}) and (\ref{homog-ind}) can be naturally extended to apply to the different
sets of devices, yielding
global parametrizations of the aforementioned manifolds
${\cal C}_r$, ${\cal C}_c$ and ${\cal C}_l$. 
In the resistive case we may write
%The global parametrization of ${\cal C}_r$ can be written as 
\ifthenelse{\boolean{ieee}}{}{\vspace{-5mm}}
\begin{subequations} \label{homog-res2}
  \begin{eqnarray}
    i_r & = & \psi_r(u_r) \\
    v_r & = & \zeta_r(u_r),
  \end{eqnarray}
\end{subequations}
the $k$-th entries of $\psi_r$ and $\zeta_r$ defining the parametrization (\ref{homog-res})
of the 
$k$-th resistor. %In (\ref{homog-res2}), t
The $m_r$-dimensional homogeneous variable $u_r$ 
lies on the space
%\begin{eqnarray} %\label{Hr}
$\mathbb{H}_r = \R^{r_1} \times \mathbb{T}^{r_2},$
%\end{eqnarray}
%and 
with $r_1+r_2=m_r$. The first factor in $\mathbb{H}_r$ %(cf.\ (\ref{Hr}))
accommodates the parametrization domains for resistors whose characteristics are not closed 
curves, %so that each such characteristic is 
each one of them being therefore diffeomorphic, under our working assumptions, 
to the real line $\R$
(w.l.o.g.\ we order the resistive branches in a way such that these are the first
%$m_{r_1}$ 
ones).
In turn, $\mathbb{T}^{r_2}$ denotes
the torus $\mathbb{S}^1 \times \stackrel{(r_2)}{\ldots} \times \mathbb{S}^1$ and
defines the domain of the homogeneous description of the set of resistors whose characteristics define
 closed curves (to be termed 
{\em loops} in the sequel).  %(and $r+s=m$). 
In the absence of loops 
$\mathbb{H}_r$ amounts to $\R^{m_r}$; this is very often the case 
in circuit theory and it is always met in the 
linear setting. %In general, 
%[somewhere mention loops/cycles in the characteristics]. 
Note also that both $\psi_r$ and $\zeta_r$ are
smooth maps $\mathbb{H}_r \to \R^{m_r},$
and that the manifold ${\cal C}_r$ %\subseteq \R^{2m_r}$ 
accommodating the characteristics of all resistors  
is the image of the map $(\psi_r, \zeta_r):\mathbb{H}_r \to \R^{2m_r}$,
%this map 
which provides %providing
a global
parametrization of ${\cal C}_r$.

%in (\ref{homog-res}),
%(\ref{homog-cap}) and (\ref{homog-ind}), simply in the understanding
%that $\zeta_r$ and $\psi_r$ are now maps defined 
%between certain $m_r$-dimensional spaces (specified below)
%% $\R^{m_r} \to \R^{m_r}$
%and that, analogously, $\psi_c$, $\zeta_c$ and $\psi_l$, $\zeta_r$ have
%respectively $m_c$- and $m_l$-dimensional domains and co-domains.
%maps $\R^{m_c} \to \R^{m_c}$
%and $\R^{m_l} \to \R^{m_l}$.
%Specifically, 

Analogously,
%for reactive devices we have
%DC voltage sources: $i_v=u_v$ ``free'' variable, $v_v=V$.
%DC current sources: $v_j = u_j$ ``free'' variables, $i_j=I$.
%global parametrizations of the form
%%\begin{subequations}
%\begin{eqnarray} \label{cap-param}
%\sigma_c = \psi_c(u_c), \ v_c = \zeta_c(u_c)
%\end{eqnarray}
%%\end{subequations}
%for capacitors, and
%%\begin{subequations}
%\begin{eqnarray} \label{ind-param}
%i_l = \psi_l(u_l), \  \varphi_l=\zeta_l(u_l)
%\end{eqnarray}
%%\end{subequations}
%for inductors. As before, 
the reactive homogeneous variables $u_c$ and $u_l$ lie
on %the spaces 
$\mathbb{H}_c = \R^{c_1} \times \mathbb{T}^{c_2} \ \text{ and } \
\mathbb{H}_l = \R^{l_1} \times \mathbb{T}^{l_2},$ respectively,
with the same splitting of variables in both cases. 
For capacitors, we get a global parametrization of ${\cal C}_c$ by joining together
the parametrizations (\ref{homog-cap}) of the individual devices to get
\begin{subequations} \label{homog-cap2}
\begin{eqnarray}
\sigma_c & = & \psi_c(u_c) \\ v_c & = & \zeta_c(u_c),
\end{eqnarray}
\end{subequations}
and the same goes for inductors, for which the individual parametrizations (\ref{homog-ind}) define
the maps
\begin{subequations} \label{homog-ind2}
\begin{eqnarray} 
i_l & = & \psi_l(u_l) \\  \varphi_l & = & \zeta_l(u_l).
\end{eqnarray}
\end{subequations}
As before, $\psi_c$ and $\zeta_c$ are
smooth maps $\mathbb{H}_c \to \R^{m_c}$ and, analogously, 
 $\psi_l$ and $\zeta_l$ are
maps $\mathbb{H}_l \to \R^{m_l}$. We are denoting by $m_c$ and $m_l$ the number of capacitors
and inductors, respectively, with $m_c=c_1+c_2$, $m_l=l_1+l_2$. Note also that
the manifolds ${\cal C}_c$ and ${\cal C}_l$ are the images
of the maps $(\psi_c, \zeta_c):\mathbb{H}_c \to \R^{2m_c}$ and $(\psi_l, \zeta_l):\mathbb{H}_l \to \R^{2m_l}$.  
%*Esto desubicado, quiza con individual devices... We refer the reader to Fig.\
%%\ref{fig-vdpsingcross}
%1(a) for an example of a characteristic defining a closed curve,
%standing in this case for a hysteresis loop in a nonlinear inductor.

%We finally remark that coupling effects may be included in (\ref{homog-res2}),
%(\ref{homog-cap2}) and (\ref{homog-ind2}) by assuming a (say) non-diagonal dependence
%of the different functions on their arguments. The description of such effects is however 
%more complicated in general since the aforementioned classification theorem  of 1-manifolds
%cannot be used;... ****

%By inserting the parametrizations (\ref{res-param}), (\ref{cap-param}) and (\ref{ind-param})
%into the general model (\ref{general}), we get\hspace{-1mm}
\ifthenelse{\boolean{ieee}}{{\it Kirchhoff laws and homogeneous model.}}{\paragraph{Kirchhoff laws and homogeneous model.}}
In order to derive the full homogeneous model we need to add
the electromagnetic relations $\sigma_c'  =  i_c, \  \varphi_l'  =  v_l,$ and also
Kirchhoff laws. These can be written as $Ai=0, \ Bv=0,$ where $i$ and $v$ denote
the $m$-dimensional vectors of currents and voltages
(with $m=m_c + m_l + m_r$ denoting the total number of branches), whereas $A$ and $B$ are 
reduced cut and cycle matrices
(find details e.g.\ in \cite{bollobas, wsbook, homoglin}). By splitting these matrices and,
as before, the current/voltage
vectors in terms of the capacitive, inductive or resistive nature of the circuit devices, 
Kirchhoff laws read as $A_c i_c + A_l i_l + A_r i_r = 0$ and $B_c v_c + B_l v_l + B_r v_r = 0,$
respectively. 

Altogether, these relations and the parametrizations (\ref{homog-res2}), (\ref{homog-cap2})
and (\ref{homog-ind2})
%understood to be defined on the domains $\mathbb{H}_r$, $\mathbb{H}_c$ and $\mathbb{H}_l$ defined above, 
make it
possible to write the equations of any uncoupled, smooth, possibly nonlinear RLC circuit as
\begin{subequations} \label{intermediate}
\begin{eqnarray}
  \psi_c'(u_c) u_c' & = & i_c \label{intermediatea}\\
  \zeta_l'(u_l)u_l' & = & v_l \label{intermediateb}\\
0 & = & A_c i_c + A_l\psi_l(u_l)   + A_r\psi_r(u_r) \label{intermediatec} \\
0 & = &B_c\zeta_c(u_c) + B_lv_l + B_r\zeta_r(u_r).\label{intermediated}
\end{eqnarray}
\end{subequations}
We may further eliminate the variables $i_c$ and $v_l$
by means of the first two equations, to get the homogeneous model
% inserting (\ref{intermediatea}) and (\ref{intermediateb})
%into (\ref{intermediatec})-(\ref{intermediated}),
\begin{subequations}\label{core}
\begin{eqnarray}
A_c \psi_c'(u_c) u_c' + A_l \psi_l(u_l) + A_r\psi_r(u_r) & = & 0 \label{corea} \\
B_c\zeta_c(u_c) + B_l\zeta_l'(u_l)u_l' + B_r\zeta_r(u_r) & = & 0. \label{coreb}
\end{eqnarray}
\end{subequations}
%or
%\begin{subequations}
%\begin{eqnarray}
%A_c (\psi_c(u_c))' + A_l \psi_l(u_l) + A_r\psi_r(u_r) & = & 0 \\
%B_c\zeta_c(u_c) + B_l(\zeta_l(u_l))' + B_r\zeta_r(u_r) & = & 0 
%\end{eqnarray}
%\end{subequations}
This approach yields a description of the circuit dynamics on the
$m$-dimensional homogeneous space $\mathbb{H}=\mathbb{H}_c \times \mathbb{H}_l \times \mathbb{H}_r$
where the homogeneous variables $u = (u_c, u_l, u_r)$ lie. 
We emphasize that
only one variable per branch is involved in the model 
%(by contrast to the tableau approach; see e.g.\ \cite{chuadesoerkuh})
but, at the same time (as far as all devices are assumed to be smooth and uncoupled),
there is no loss of generality in the formulation of this reduced
model.
%performing the reduction from the general model (\ref{general}) to (\ref{core}). 
The compactness and generality of
(\ref{core}) makes it suitable for different analytical purposes and we will exploit this
in Section \ref{sec-state}. Remember that the values of the classical circuit variables are obtained from the solutions
of this model via 
%(\ref{homog-res}), (\ref{homog-cap})
%and (\ref{homog-ind})... or... by means of the relations $i_r = \psi_r(u_r)$, 
%with $\psi_r:\mathbb{H}_r \to \R^{m_r}$, $v_r = \zeta_r(u_r)$, etc. 
%NO: USAR 
(\ref{homog-res2}), (\ref{homog-cap2})
and (\ref{homog-ind2}).
%later sections.
%\begin{boxedminipage}{16.5cm}
%\
%(e.g. state-space/impasse problem... Equilibria...)

Also worth recalling is the fact that this model encompasses in particular classical ones (formulated in 
terms of currents, voltages, charges and/or fluxes), which are simply obtained by choosing
appropriately the $\psi$ and $\zeta$ maps (e.g.\ if all resistors are assumed to be voltage-controlled
we simply fix $\zeta_r=\mathrm{id}$, so that $u_r = v_r$ and $\psi_r$ amounts to the voltage-to-current
characteristic). With this in mind, (\ref{core}) provides a general model where
all possible controlling relations can be accommodated. A simple example illustrating this, in the memristive
context,
can be found in subsection \ref{subsec-exmem}.

%\newpage
%\subsection{
\ifthenelse{\boolean{ieee}}{{\it Homogeneous variables and the homogeneous space.}}{\paragraph{Homogeneous variables and the homogeneous space.}}
%\label{subsec-homogvar}
%We finish this section with a brief remark on the nature of homogeneous variables. 
The proof of 
the classification theorem of 1-manifolds (cf.\ %Milnor's book 
\cite{milnor})
%referred to in subsection \ref{subsec-curves}  
makes use of the arc-length to build the global parametrization 
%%%%%%%%%%$g$ 
%%%%%%%%%% this remains to be updated in the  version upload to ArXiv
$\Gamma$
mentioned
in subsection \ref{subsec-curves} above;
it is then possible, after fixing a distinguished point and an orientation 
in each individual
characteristic, %curve,  
to think of the corresponding 
scalar variable $u$ as the arc-length of the curve, setting $u=0$ for that
distinguished point and defining positive/negative values of $u$ accordingly to the
chosen orientation. But there is not really a need to privilege this particular
choice; in fact, the map $\Gamma$, %referred to there 
and the
variable $u$ itself, are defined only up to a 
diffeomorphism %(automorphism?) 
of $\R$ or $\mathbb{S}^1$, respectively. This is analogous to what happens in the
linear case, where
$u$ is defined only up to a (linear) isomorphism of $\R$ (cf.\ \cite{homoglin}). 
%By using the chain rule, the reader
%can easily check that
%the pair of homogeneous coordinates characterizing the incremental homogeneous resistance
%in Proposition ** does
%not depend on this diffeomorphism, and that the same applies for capacitors and inductors.

This similarity with the linear case supports calling 
$u$ a {\em homogeneous variable} also in the nonlinear
setting, and %for this reason 
we extend the use of the term to call
$\mathbb{H}=\mathbb{H}_c \times \mathbb{H}_l \times \mathbb{H}_r$ 
the {\em homogeneous space}. By construction, this space
is diffeomorphic to the manifold ${\cal C}_c \times {\cal C}_l \times {\cal C}_r$
which accommodates the characteristics of all devices. %as described in subsection \ref{subsec-main},
%and actually provides a convenient way to handle this manifold in practice. %broadly general terms.

%(and the map --not easy letter, $h=h_c \times h_r \times h_l$ maybe-- is a global diffeomorphism).
%coming from ``homogeneous''
%with $\mathbb{H}_c = \R^{c_1} \times \mathbb{T}^{c_2}$ and
%$\mathbb{H}_l = \R^{l_1} \times \mathbb{T}^{l_2}$ etc
%\subsection{Parametric description of nonlinear electrical devices}

\subsection{Example: Van der Pol's system with a closed characteristic in the inductor}
\label{subsec-vdp}

We show in what follows how the models
above can be used in practice, focusing on a low-scale example. In particular
we will illustrate how the homogeneous model (\ref{core}) 
naturally accommodates
trajectories evolving on regions where
classical (current/voltage, or even charge/flux) descriptions do not hold globally, whereas 
homogeneous ones do; this way we
avoid the need to resort to piecewise descriptions of the reduced dynamics. We also illustrate
how partially homogeneous models, combining classical variables with homogeneous ones, provide
a useful simplification in practice, based on the fact that for many devices 
a global description in terms of one of the classical variables %(voltage, current, flux or charge) 
is often justified by physical reasons. %loses no generality.

%We illustrate the approach above
%by examining the form that the homogeneous model \eqref{core} takes for 
To this end, consider the well-known Van der Pol system,
defined by a (parallel, in the present case and without loss of generality) %series 
connection of a capacitor, an inductor and a resistor.
%, and
%assuming that all three devices are nonlinear (but
%smooth and regular). 
An admissible choice for the reduced cut and cycle matrices is
\ifthenelse{\boolean{ieee}}{$$A= \begin{pmatrix} A_c & A_l & A_r \end{pmatrix} =\begin{pmatrix} 1 & -1 & 1 \end{pmatrix},$$
$$B= \begin{pmatrix} B_c & B_l & B_r \end{pmatrix} =
\begin{pmatrix} 1 & 1 & 0 \\ 1 & 0 & -1 \end{pmatrix}.$$}
{$$A= \begin{pmatrix} A_c & A_l & A_r \end{pmatrix} =\begin{pmatrix} 1 & -1 & 1 \end{pmatrix}, \ B= \begin{pmatrix} B_c & B_l & B_r \end{pmatrix} =
\begin{pmatrix} 1 & 1 & 0 \\ 1 & 0 & -1 \end{pmatrix}.$$}
%and using (...)-(...)-(...), the system \eqref{core} reads as
%\begin{eqnarray} \label{vdp}
%p(u_c)u_c' & = & \psi_l(u_l) \\
%q(u_l) u_l' & = & -\zeta_c(u_c) - \zeta_r(u_r) \\
%0 & = & -\psi_l(u_l) + \psi_r(u_r),
%\end{eqnarray}
%Series:
%$$A= \begin{pmatrix} A_c & A_l & A_r \end{pmatrix} =\begin{pmatrix} -1 & 1 & 0 \\ 0 & -1 & 1 \end{pmatrix}, \
%B= \begin{pmatrix} B_c & B_l & B_r \end{pmatrix} =
%\begin{pmatrix} 1 & 1 & 1 \end{pmatrix}$$
%and using (...)-(...)-(...), the system \eqref{core} reads as
%\begin{eqnarray} \label{vdp}
%p(u_c)u_c' & = & \psi_l(u_l) \\
%q(u_l) u_l' & = & -\zeta_c(u_c) - \zeta_r(u_r) \\
%0 & = & -\psi_l(u_l) + \psi_r(u_r),
%\end{eqnarray}
%with $p(u_c)=\psi_c'(u_c),$ $q(u_l)=\zeta_l'(u_l)$. We write the constraint as the last equation for better c%larity.
%Series - END

\noindent If we avoid imposing a specific control variable
for each device (that is, if 
the resistor is not assumed to be either current-controlled 
or voltage-controlled,
etc.) we get a completely general model of the Van der Pol circuit dynamics
in terms of homogeneous variables $u_c$, $u_l$, $u_r$, which are scalar in this example
since there is exactly one device of each type.
%, suitable  not only for classical qualitative studies but also for the analysis of impasse points
%in certain scenarios. 
This is made possible by the global parametric descriptions
%of the forms 
(\ref{homog-res}), (\ref{homog-cap}) and (\ref{homog-ind}). With the above choice
for $A$, $B$, the model (\ref{core}) reads %for our example 
as
\begin{subequations}\label{vdp}
\begin{eqnarray} 
%p(u_c)
\psi_c'(u_c)u_c' & = & \psi_l(u_l) - \psi_r(u_r)\\
%q(u_l) 
\zeta_l'(u_l)u_l' & = & -\zeta_c(u_c) \\
0 & = & \zeta_c(u_c) - \zeta_r(u_r).
\end{eqnarray}
\end{subequations}
%with $p(u_c)=\psi_c'(u_c),$ $q(u_l)=\zeta_l'(u_l)$

%so that the model above accommodates all possible settings (always under the hypothesis that all characteristics
%are smooth --somewhere: indicate we use this term, when referring to a set, that it is a manifold. Particular excludes
%self-intersections). MAYBE SUBSECTION PARTIALLY HOMOGENEOUS MODELS... 
%Should e.g.\ the capacitor and the inductor be linear (with capacitance $C$ and inductance $L$)
%and the resistor current-controlled, this would correspond to setting $u_c=v_c$ (i.e.\ $\zeta_c=\mathrm{id}$),
%$\psi_c(u_c)=Cu_c$; $u_l =i_l$ (that is, $\psi_l=\mathrm{id}$), $\zeta_l(u_l)=Lu_l$, ***id r. 
%If, on the contrary, one wants to model impasse points
%under the assumption that the resistor is voltage-controlled by a global
%function $g(v_r)$, simply set $u_r=v_r$ (i.e.\ $\zeta_r=\mathrm{id}$),
%$\psi_r=g$.
%Generality (\ref{vdp}) without assuming any global description. Ref reactive later (or here). 

%\vvv

%\noindent {\bf Van der Pol's system with under a hysteresis-like loop in the inductance.} 
\noindent Several simplified versions of this model will be derived for different purposes,
%and 
under specific %certain 
assumptions on the devices.
%To be specific, let us 
Assume first %for instance % on the dynamics of the circuit defined by the parallel connection of
the capacitor to be linear and voltage-controlled:
the variable $u_c$ can be then %simply 
taken
to be $v_c$ (that is, $\zeta_c$ amounts to the identity), 
with $\psi_c(v_c)=Cv_c$, $C$ being the capacitance.
This yields a
{\em partially homogeneous} model, namely\hspace{-2mm}
\begin{subequations}\label{vdp1}
\begin{eqnarray} 
%p(u_c)
Cv_c' & = & \psi_l(u_l) - \psi_r(u_r)\\
%q(u_l) 
\zeta_l'(u_l)u_l' & = & -v_c \\
0 & = & v_c - \zeta_r(u_r).\label{vdp1c}
\end{eqnarray}
\end{subequations}
Additionally, the resistor will be assumed to be voltage-controlled
%, a voltage-controlled resistor 
%with a characteristic 
by a relation of the form
$i_r %=g(v_r)
=-v_r + v_r^3$, as in the parallel version of the classical Van der Pol system
(which would be obtained after an additional linear assumption on the inductor). 
This implies that we may further 
take $u_r$ to be the voltage
$v_r$ (equivalently, $\zeta_r$ amounts to the identity), with
$\psi_r(v_r)%=g(v_r)
=-v_r + v_r^3$. This results in
\begin{subequations}\label{vdp2}
\begin{eqnarray} 
Cv_c' & = & \psi_l(u_l) - \psi_r(v_c)\\
\zeta_l'(u_l)u_l' & = & -v_c, 
\end{eqnarray}
\end{subequations}
where we have eliminated $v_r$ in light of the identity $v_r=v_c$.

In what follows,
the characteristic of the nonlinear inductor %whose characteristic
will be assumed to be defined by a {\em closed} %, regular planar 
curve, an assumption which makes it convenient to
%as detailed later  makes the 
keep a homogeneous description 
%formalism convenient 
for this device.
Specifically, the current-flux relation
will be assumed to lie on the curve depicted in
Fig.\ 1(a). 
%(current and flux in the $x$ and $y$-axes, respectively). 
%\ref{fig-vdpsingcross}(a). Ref. does not work (?)
Such loops typically arise in the presence of hysteresis
phenomena (see e.g.\ \cite{lacerda}, where a Jiles-Atherton model for ferroresonance %phenomena
in a ferromagnetic core yields a loop such as the one displayed in the figure). 
%The effect of the presence of a nonlinear
%inductance in Van der Pol's circuit has been examined e.g.\ in \cite{appelbe2,
%bartucelli}. 
We 
give the loop a parametric description following 
%Model, from Lapshin 2017, arxiv, 
\cite{lapshin}, namely %specifically in the form
\begin{subequations} \label{lapshin-eqs}
\begin{eqnarray} 
%i_l & = &a(\cos(u))^m + b_x(\sin(u+\Delta))^n \\
%\varphi_l & = & b_y\sin(u)
i_l & = & \psi_l(u_l) = \alpha\cos^m\hspace{-0.3mm}u_l + \beta \sin^n(u_l+\delta) \\
\varphi_l & = & \zeta_l(u_l) = \gamma \sin u_l,
\end{eqnarray}
\end{subequations}
for certain parameters $m$, $n$, $\alpha$, $\beta$, $\gamma$ and $\delta$.
Set %Fix the values 
$m=n=3$, $\alpha=0.2$, %formerly a
$\beta=\gamma=1$, %formerly  $b_x=b_y=1$, 
$\delta=0.05$.

Our goal is simply to illustrate the convenience of using
a model such as (\ref{vdp2}) to track trajectories along which a global
current- or flux-controlled description of the inductor does not apply, because of the closed
nature of the characteristic governing the nonlinear inductor.
Note, indeed, that at local extrema of the curve in Fig.\ 1(a)
%To fix some elementary terms, note that this curve exhibits two extrema 
(where the flux meets local maxima or minima) 
%and minimum values, respectively, and which display a horizontal tangent)
we have $\zeta_l'(u_l) = 0$ and near such points there is 
no local flux-controlled description of the characteristic.
Similarly, at turning points 
%and  several (elaborate: more than two depending on a parameter***: at least those with extremal values
%for the current) turning points 
(points with a vertical tangent) we have $\psi_l'(u_l)=0$ and
there is no local current-controlled 
description of the curve.
%Extrema and turning points will be jointly called {\em critical points}. 
In order to describe the dynamics of the circuit
in a given region in terms of a state-space model, the flux would be 
precluded as a model variable for trajectories which reach
at least one of the aforementioned extrema and, analogously, 
the inductor current would be ruled out for trajectories
%reaching (one or more) 
undergoing turning points. 
%[MAYBE DRIVE THE FOLLOWING SENTENCE TO THE
%SINGULARITY-CROSSING PARAGRAPH BELOW. JUST KEEP SOMETHING BRIEFER Our goal is 
%however to formulate a global model in terms of a single state variable in the inductor] 
Obviously, there is no chance to formulate a single state model in terms of either
the flux or the current if we want such a model to cover trajectories reaching both
extrema {\em and} turning points. Such a trajectory, 
stemming from the initial point %defined by $v_0=
$(0.500, -1.805)$ %8051$ 
and approaching a limit cycle,
is depicted in Fig.\ 1(b); %computer simulations show that 
a zero of $\zeta_l'(u_l)$ %q
is met at the point $(0,-1.571)$ for $t=0.100$. %a sequence of
Zeroes of $\psi'_l(u_l)$  %p
are found  at
the points $(0.124, -1.621)$,
%[Maple v(t) = 0.124754326041131, 
%            u(t) = -1.62128213670003]
$(-0.883, 0.043)$,
%dsn(1.106); #1.206 en el artículo por el +0.100
%[t = 1.106, v(t) = -0.882604924933317, u(t) = 0.0430433563775830]
$(-0.882, 0.058)$,
%dsn(1.123); #1.223 en el artículo
%[t = 1.123, v(t) = -0.882060047266184, u(t) = 0.0580622157311214]
$(1.089, 0.059)$,
%dsn(3.047); #3.147 en el artículo
%[t = 3.047, v(t) = 1.08879504059291, u(t) = 0.0590485592806756]
$(1.088,0.043)$, etc.,
%dsn(3.061); #3.161 en el artículo
%[t = 3.061, v(t) = 1.08865430164475, u(t) = 0.0437861319994231]
for $t=0.080$, $1.206$, $1.223$, $3.147$, $3.161$,
and so on.
%2064$, $1.2232$, $3.1476$, $3.1614$, etc.
%where extrema are met at ... and turning points at ... 
%Note that the trajectory approaches a limit cycle, as in the classical Van der Pol oscillator.
The fact that (\ref{vdp2}) holds globally is the key for the model to accommodate 
such trajectories.
If needed, the values of the current $i_l$ and $\varphi_l$ along the trajectory
can be explicitly computed via (\ref{lapshin-eqs}). 
\ifthenelse{\boolean{ieee}}{\begin{figure}[ht] \label{fig-vdpsingcross}
%\begin{center}
\vspace{2mm}
\hspace{-2mm} 
\parbox{0.5in}{%\hspace{-1mm}
\epsfig{figure=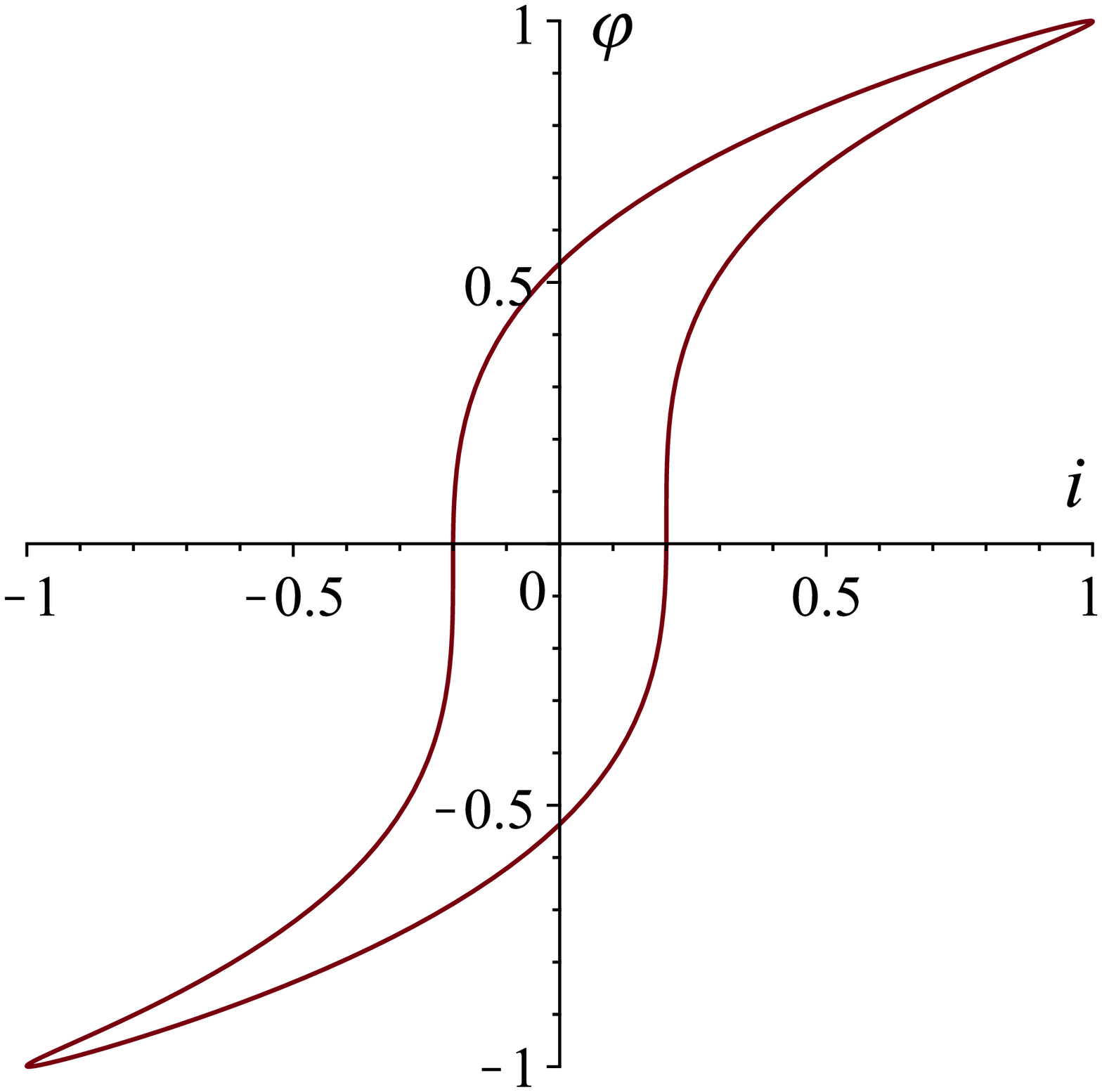, width=0.23\textwidth} %This name: pdf obtained from converting the .eps%\epsfig{figure=hysteresis--final, width=0.23\textwidth} %This one: pdf directly exported from Maple. Sizes are different
%\epsfig{figure=hys_loop_letters, width=0.23\textwidth} %This one has the labels placed ``in the middle'' of the axes (Maple default)
%\epsfig{figure=hysteresis_loop, width=0.23\textwidth} %the same figure as hys..a0.2... below, just simplified the name
%\epsfig{figure=hysteresis_loop_a0.2_delta0.05.eps, width=0.23\textwidth}
%\put(-2,70){{\footnotesize $i_l$}}
%\put(-76,148){{\footnotesize $\varphi_l$}}
}
\hspace{35mm}
\parbox{0.5in}{%\hspace{-1mm}
\epsfig{figure=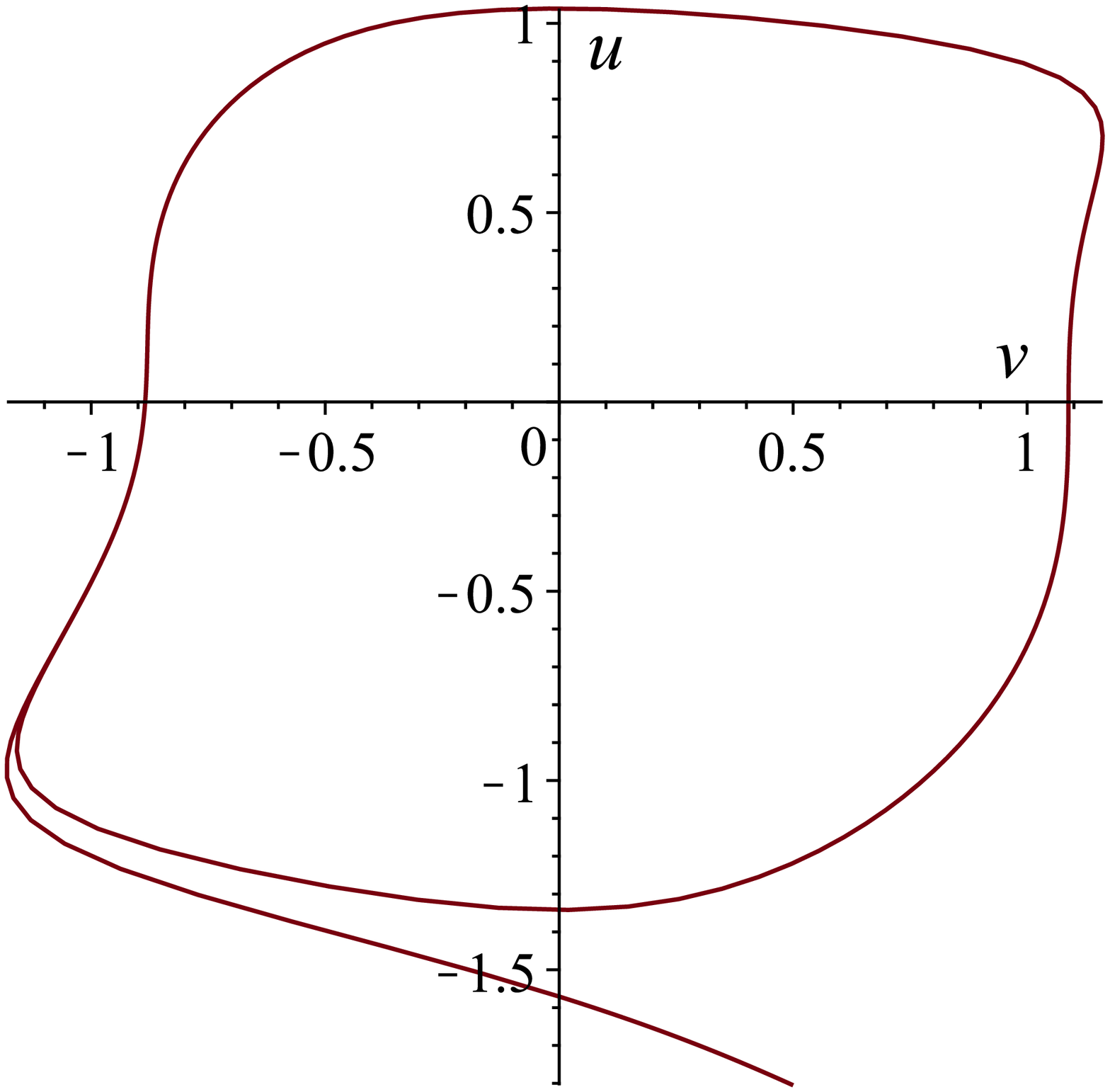, width=0.23\textwidth}}
\caption{\noindent (a) Hysteresis loop (\ref{lapshin-eqs}) in the inductor of %a modified
Van der Pol's circuit (abscissae: $i_l$, ordinates: $\varphi_l$). \ \ %\newline 
(b) A trajectory of (\ref{vdp2}) %approaching a limit cycle and 
undergoing
both turning points and extrema of the %hysteresis 
loop
(abscissae: $v_c$, ordinates: $u_l$).}
%\end{center}
\end{figure}}{\begin{figure}[ht] \label{fig-vdpsingcross}
%\begin{center}
\vspace{2mm}
\hspace{12mm} 
\parbox{0.5in}{%\hspace{-1mm}
\epsfig{figure=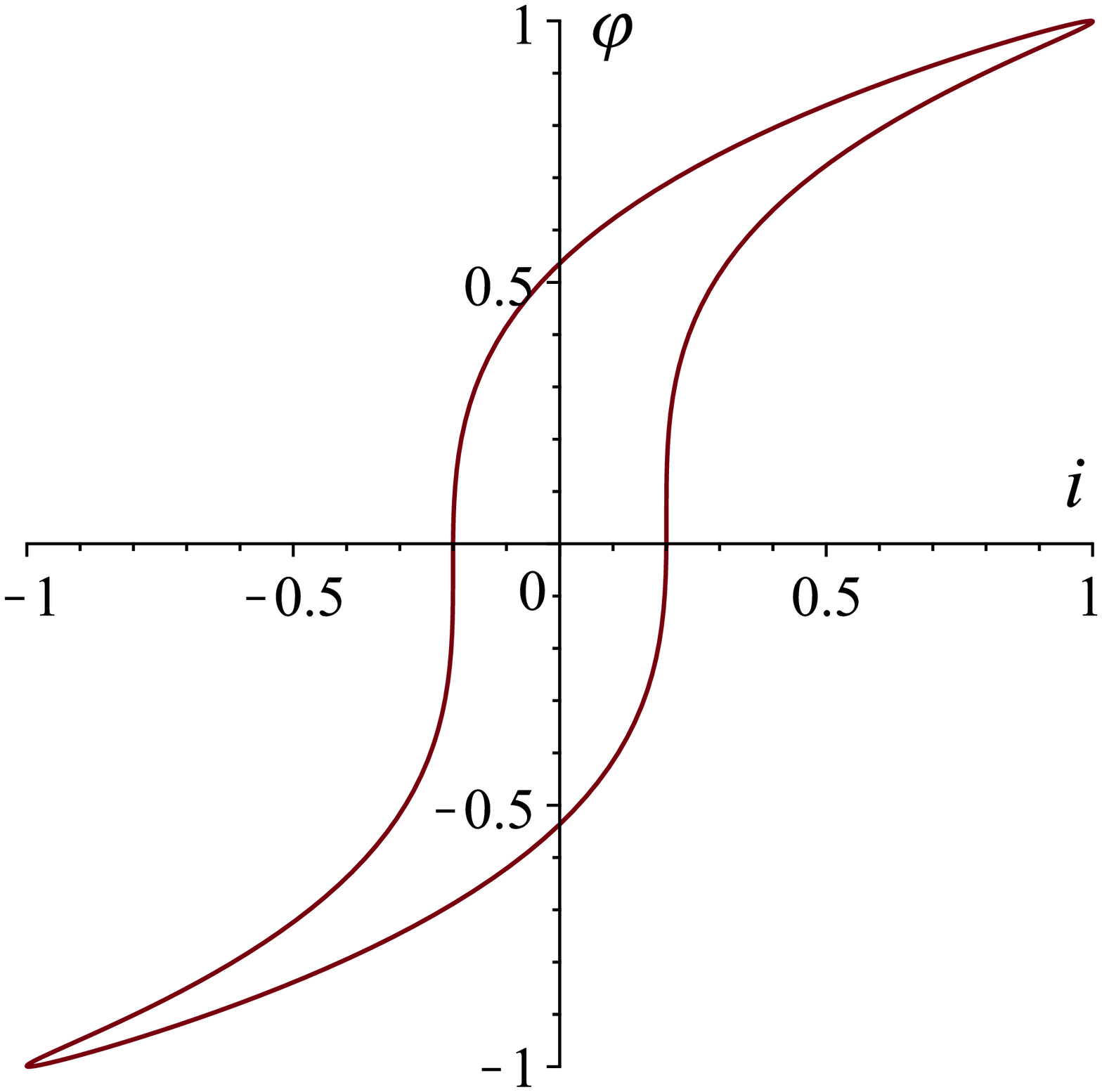, width=0.35\textwidth}  %hysteresis_loop_a0.2_delta0.05, width=0.35\textwidth}
%\put(0,70){{\footnotesize $i_l$}}
%\put(-76,148){{\footnotesize $\varphi_l$}}
}
\hspace{67mm}
\parbox{0.5in}{%\hspace{-1mm}
\epsfig{figure=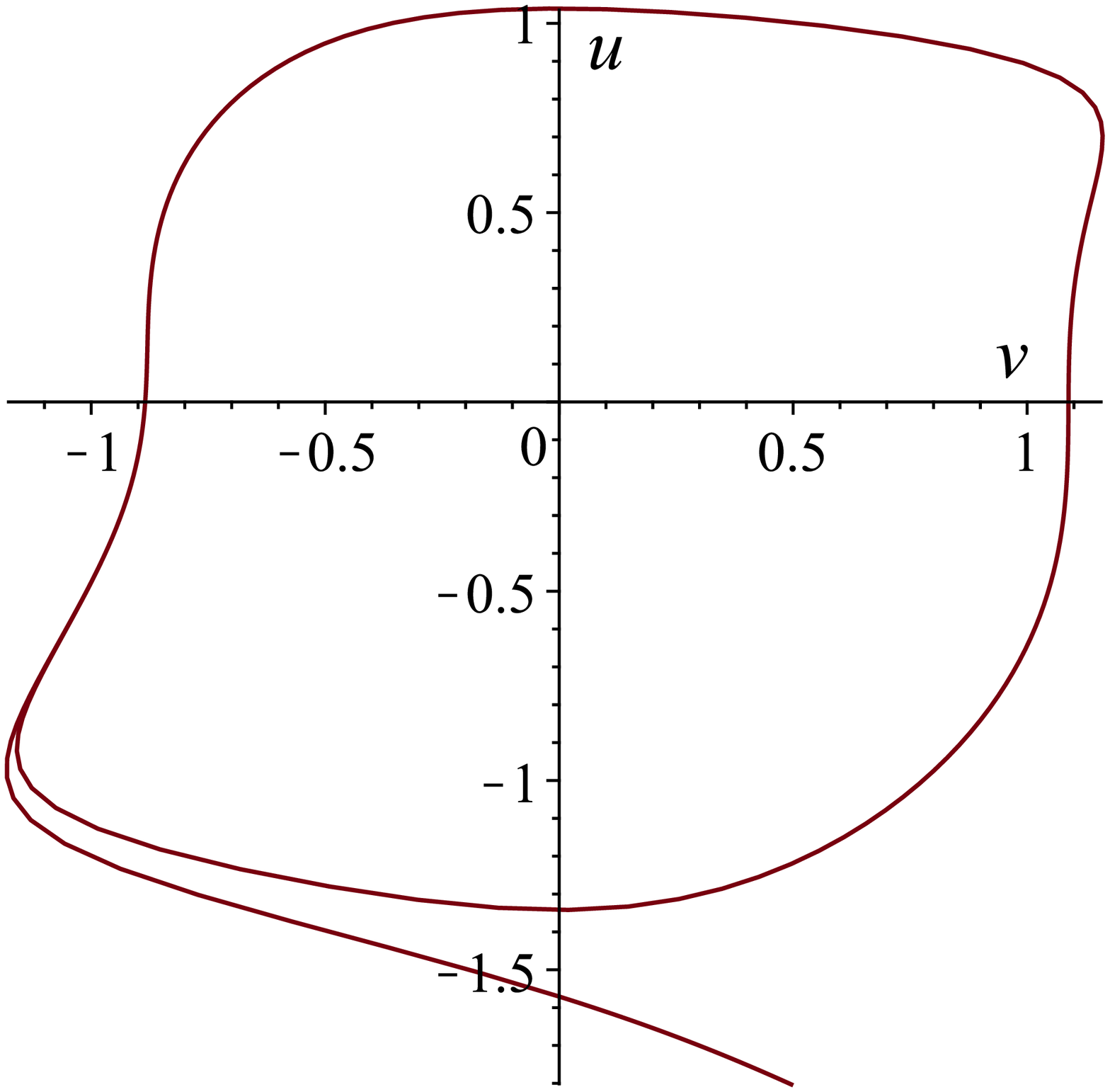, width=0.35\textwidth}} %   _NoLetters, width=0.35\textwidth}}
\caption{\noindent (a) Hysteresis loop (\ref{lapshin-eqs}) in the inductor of %a modified
Van der Pol's circuit (abscissae: $i_l$, ordinates: $\varphi_l$). \ \ %\newline 
(b) A trajectory of (\ref{vdp2}) %approaching a limit cycle and 
undergoing
both turning points and extrema of the %hysteresis 
loop
(abscissae: $v_c$, ordinates: $u_l$).}
%\end{center}
\end{figure}
}

%Notation: $u_l$ for the inductor, $v_c$ for the capacitor

%\color{red}
%\subsection{A reactive example...VdPol pdte}

%On the other Van der Pol - integration with a hysteresis in the inductance, global approach

%\color{black}

\subsection{Controlled sources and coupled problems}
\label{subsec-coupling}

The essential ideas behind the homogeneous formalism can be extended 
to circuits including controlled sources and coupling effects. Even 
if, generally speaking, there is no higher-dimensional analog to the classification theorem
of 1-manifolds, 
%(that is, there is no chance to guarantee the general 
%existence of global parametrizations of manifolds with dimension two or higher),
%virtually all 
%the vast majority of
most
cases of interest may in practice be described in terms of
homogeneous parameters. For the sake of brevity, we just illustrate in what follows how
this is possible for controlled sources and for linearly coupled devices.

For controlled sources, we extend the ideas introduced in \cite{homoglin}
by considering two circuit branches governed by
\begin{subequations}\label{controlled}
\begin{eqnarray}
f_1(i_1, v_1) & = & 0 \\
p_2 v_2 -q_2 i_2 + f_2(i_1, v_1) & = & 0. \label{controlled2}
\end{eqnarray}
\end{subequations}
The first equation describes the controlling branch,
which is assumed to be a nonlinear resistor.
The controlled source is governed by the latter equation.
By means of the
parameters $p_2$ and $q_2$, which are required not to vanish
simultaneously, we include both (controlled) voltage and 
current sources in the same framework; note also that there is
no need to restrict the analysis to cases in which
only one specific variable (current or voltage) controls the source. 

It is clear that the 
controlling device admits a homogeneous description of the form $i_1=\psi_1(u_1)$,
$v_1 = \zeta_1(u_1)$. For the controlled source, set
\begin{subequations} \label{controlled-bis}
\begin{eqnarray}
i_2  =  \psi_2(u_1, u_2) = p_2 u_2 + \frac{q_2}{p_2^2+q_2^2}f_2(\psi_1(u_1), \zeta_1(u_1))  \ \ \ \\
v_2  = \zeta_2(u_1,u_2)=  q_2 u_2 - \frac{p_2}{p_2^2+q_2^2}f_2(\psi_1(u_1), \zeta_1(u_1)).  \ \ \ 
\end{eqnarray}
\end{subequations}
Altogether, the maps $\psi=(\psi_1, \psi_2)$ and $\zeta=(\zeta_1, \zeta_2)$
provide a global parametrization of the %controlling-controlled
characteristics (\ref{controlled}), describing the controlled source
and its controlling device in terms
of the homogeneous variables $(u_1, u_2)$. It should be clear
that these relations can be combined with the remaining characteristics
and Kirchhoff laws to get again a system of the form (\ref{core}),
in the understanding that the maps $\psi_r$ and $\zeta_r$ there do no longer
display a diagonal structure. The reader is referred to 
\cite{homoglin} for a detailed discussion concerning
the way in which this approach
 allows, in the linear setting, for a unified treatment
of small-signal equivalents of different types of transistors,
actually involving different types of controlled sources.

%Similar remarks apply to l
Linearly coupled devices can be easily handled in homogeneous
terms. Focus, for instance,
on a pair of coupled linear inductors with self-inductances
$L_1$, $L_2$ and mutual inductance $M$. These coupled devices
admit the homogeneous description defined by
$i_1=\psi_1(u_1)=u_1$,
$i_2 = \psi_2(u_2)=u_2$ and
\begin{subequations}
\begin{eqnarray}
\ \varphi_1 =  \zeta_1(u_1, u_2) = L_1u_1 + Mu_2 \  \\
\  \varphi_2  =  \zeta_2(u_1,u_2)= M u_1 +L_2u_2. 
\end{eqnarray}
\end{subequations}
The underlying idea here, which applies 
in many other contexts, is that a plane (or actually, any 
$n$-dimensional linear subspace of $\R^m$, with $n=2$, $m=4$ in the case above)
%coupled inductors described above) 
obviously admits a global linear
parametrization. 
As before, we may use such descriptions in models of the form
(\ref{core}), with the remark that in the presence of 
coupled inductors the diagonal structure of the corresponding
maps $\psi_l$ and $\zeta_l$ is lost. 
The same 
ideas apply to linearly coupled capacitors and resistors, and
we leave details in this direction to the reader.

The cases considered above briefly indicate how the homogeneous framework
can be extended in order to accommodate coupled devices, at least
in certain scenarios. In
the same direction, the analysis of circuits including multiports and 
multiterminal elements is in the scope of future research.

%\section{Analytical properties of nonlinear circuits in the homogeneous setting}
\section{The state-space problem in the homogeneous setting}
%. passe points} %. Solution manifold, impasse, etc}
\label{sec-state}

%%%%%MAYBE REUSE SOMETHING: Our goal now is to define a reduced model [$m$-dimensional reduction]
%in terms of just three vector variables, one for each type of device. This is typically
%performed by assuming an explicit global description of each individual 
%device in terms of one of the two variables in its characteristic
%(that is, either in terms of $v_r$ or of $i_r$ for resistors, etc.). 
%Needless to say there is a loss of generality in this approach.
%We show in the sequel how to perform an $m$-dimensional reduction
%[define this somewhere] without such a loss of generality.]

The %Our 
formalism 
introduced above 
provides a framework to address in full
generality different analytical problems in circuit theory.
The key
remark is that the homogeneous space $\h = \h_c \times \h_l \times \h_r,$ where
the homogeneous variables $u$ lie, together with the %homogeneous 
model 
(\ref{core}),
provide a reduced setting for such analyses which avoids
%for any 
unnecessarily restrictive hypothesis on controlling variables.
In this section we apply such framework to a classical problem
in nonlinear circuit theory, namely the state-space reduction problem. 
We refer the reader to subsection \ref{subsec-state} for an
introduction to this problem.
%the paragraphs following (\ref{core-diff})  and (\ref{core-alg}) since these equations will allow for a simpler introduction to this problem.
%To simplify the discussion, we excludes sources from the working 
%scenario (the extension to problems
%with independent sources is straightforward), so that our model amounts to
%\begin{subequations} \label{corer}
%\begin{eqnarray}
%A_c \psi_c'(u_c)u_c' + A_l \psi_l(u_l) + A_r\psi_r(u_r) & = & 0 \\
%B_c\zeta_c(u_c) + B_l \zeta_l'(u_l)u_l' + B_r\zeta_r(u_r) & = & 0.
%\end{eqnarray}
%\end{subequations}
%Additionally, i
%In what follows w
We restrict 
the attention back to uncoupled circuits, even if many ideas can be extended to coupled
problems along the route sketched in subsection \ref{subsec-coupling} above.

Before proceeding, a brief digression on the use of the term {\em reduction} is in order.
Generically, we use this expression to mean the elimination of certain variables
from the whole set of branch currents, voltages, charges and fluxes in a given circuit.
In practice, this takes two forms in our context: on the one hand, we built in the previous
section the general
$m$-dimensional ($m$ being the total number of branches) model
(\ref{core}), involving a single variable $u$
(either $u_r$, $u_c$ or $u_l$) per branch,
instead of two (current and voltage) for resistors, or even three (current,
voltage and either charge or flux) for reactive elements: the model
(\ref{general}) below can be of help for comparison purposes. On the other,
in this section we perform a further reduction by formulating
the dynamics in terms of just $m_c+m_l$ state variables, the latter
equalling the number of reactive branches: see, specifically, (\ref{state}) but
also the quasilinear reductions briefly considered in subsection \ref{subsec-manifold}.
Needless to say, other types of reduced models arise in other circuit modelling families,
notably in nodal analysis.

\subsection{Splitting the circuit equations into
differential equations and constraints}

In order to make the 
discussion lighter we impose a restriction on the 
allowed circuit 
topologies: we assume that the circuit has neither cycles composed
exclusively of capacitors, nor cutsets composed only of inductors.
It is well known that these topological assumptions imply that the matrices
$A_c$ and $B_l$ have maximal column rank; details in this regard can be found 
in \cite{bollobas,wsbook,carentop} and references therein. Circuits satisfying this %restriction 
are
said to be {\em topologically nondegenerate}.
We also assume throughout that the circuit is
connected.

The homogeneous 
model (\ref{core}) has a differential-algebraic form. As detailed in what follows, we may rewrite
it in a way which splits the system into a set of differential
equations and a set of constraints. To do so,
denote by $m=m_c+m_l+m_r$ the total number of branches  and by
$n$ the number of nodes in the circuit. 
Let $A_c^{\perp} \in \R^{(n-1 - m_c)\times (n-1)}$ and $B_l^{\perp} \in \R^{(m-n+1 - m_l) \times (m-n+1)}$
be two full row  rank
matrices
such that $A_c^{\perp} A_c=0$, $B_l^{\perp}B_l=0$.
Allowed by the aforementioned fact that 
$A_c$ and $B_l$ have maximal column rank,
we will choose in addition 
two matrices
$A_c^- \in \R^{m_c \times (n-1)}$, $B_l^- \in \R^{m_l \times (m-n+1)}$
such that $A_c^- A_c=I_{m_c}$ and $B_l^-B_l =I_{m_l}$ (to be specific, set
$A_c^- = (A_c^{\tra}A_c)^{-1}A_c^{\tra}$, 
$B_l^- = (B_l^{\tra}B_l)^{-1}B_l^{\tra}$). By construction,
%of $A_c^{\perp}$ and $B_l^{\perp}$ we easily 
it easy to see that
\begin{eqnarray} \label{ACBL}
A_0 = \begin{pmatrix}A_c^- \vspace{0.5mm} \\ A_c^{\perp} \end{pmatrix}, \ \text{ and } \
B_0 = \begin{pmatrix}B_l^{\perp}  \vspace{0.5mm} \\ B_l^- \end{pmatrix}
\end{eqnarray}
are non-singular matrices with orders $n-1$  and $m-n+1$, respectively. 
%CAREFUL, AC- OR $A_c^{\tra}$ COMO (\ref{ACBLbis}), change there*. 
%CAREFUL ORDER B.
%In the first version I had
%B_0 = \begin{pmatrix}B_l^- \vspace{0.5mm} \\ B_l^{\perp}  \end{pmatrix}
%but in the second I change, for later simplicity (in the proof of the main theorem)

Now, by premultiplying
%each one of the two the equations in (\ref{core}) 
(\ref{corea}) by $A_0$ and (\ref{coreb}) by $B_0$ %respectively,
we get, after an obvious reordering, a splitting of the homogeneous model
into
a set of (so-called quasilinear o linearly implicit) differential equations
\begin{subequations} \label{core-diff}
\begin{eqnarray}
\psi_c'(u_c)u_c' & = & -A_c^-( A_l \psi_l(u_l) + A_r\psi_r(u_r))\\
\zeta_l'(u_l)u_l' & = & -B_l^- (B_c\zeta_c(u_c) + B_r\zeta_r(u_r))
\end{eqnarray}
\end{subequations}
and a set of constraints
\begin{subequations} \label{core-alg}
\begin{eqnarray}
A_c^{\perp} \left( A_l \psi_l(u_l) + A_r\psi_r(u_r)\right) & = & 0 \\
B_l^{\perp} \left( B_c\zeta_c(u_c) + B_r\zeta_r(u_r)\right) & = & 0.
\end{eqnarray}
\end{subequations}
%We emphasize the fact that the system defined by
%(\ref{core-diff})  and (\ref{core-alg}) is simply a recasting of (\ref{core}).

\subsection{The state-space reduction problem}
\label{subsec-state}

The circuit equations (\ref{core-diff}) and (\ref{core-alg}) will make it possible to tackle
under really broad assumptions the state-space modelling problem.
Concerning this topic, we refer the reader to classical references
% (cf.\ \cite{wsbook} and references therein).
such as \cite{chua80, chuadesoerkuh} but also to more recent approaches 
discussed e.g.\ in
\cite{wsbook, sommult, sommtree};
for memristive circuits see %and in 
\cite{corinto2016, corinto2017, membranch}
%within the context of .
and the references therein.

To present the state-space reduction problem, we drive the attention to a classical
nonlinear circuit model, namely the one obtained by writing explicitly Kirchhoff laws
and the characteristics of devices together  with the elementary
electromagnetic laws relating capacitor charges and currents, 
and inductor fluxes and voltages. This yields
\begin{subequations} \label{general}
\begin{eqnarray}
  \sigma_c' & = & i_c \\
  \varphi_l' & = & v_l \\
0 & = & A_c i_c + A_l i_l + A_ri_r  \\
0 & = &B_cv_c + B_lv_l + B_rv_r  \\ 
0 & = & f_c(\sigma_c, v_c) \label{generale}\\
0 & = & f_l(\varphi_l, i_l) \label{generalf} \\
0 & = & f_{r}(i_{r}, v_{r}).\label{generalg}
\end{eqnarray}
\end{subequations}
%where the 
%$g_c(\sigma_c, v_c)$ [], $g_l(\varphi_l, i_l)$  [] and
%$g_r:\R^{2m_r} \to \R^{m_r}$ describe the set of characteristics of capacitors,% inductors and resistors.
In the circuit-theoretic literature it is very common to impose
assumptions on the controlling variables within the characteristics
(\ref{generale}), (\ref{generalf}) and (\ref{generalg}). Say, for example,
that inductors are globally current-controlled in the form $\varphi_l=\gamma_l(i_l)$,
and capacitors and resistors globally voltage-controlled by certain maps
$\sigma_c=\xi_c(v_c)$,  and $i_r=\xi_r(v_r)$. This yields, from (\ref{general})
and again under a smoothness assumption on the reactive devices,
a reduced model of the form
\begin{subequations} \label{red01}
\begin{eqnarray}
A_c \xi_c'(v_c)v_c' + A_l i_l + A_r\xi_r(v_r) & = & 0 \\
B_cv_c + B_l\gamma_l'(i_l)i_l' + B_rv_r  & = & 0.
\end{eqnarray}
\end{subequations}
Now the problem is how to formulate conditions both on the topology of the
circuit and on the characteristics making it possible to derive, from (\ref{red01}),
a state-space model of the circuit equations, that is, a system of 
explicit ordinary 
differential equations capturing all the dynamics of (\ref{red01}) (and 
thereby
of (\ref{general})). The goal is, essentially, to eliminate $v_r$ from (\ref{red01}) to
get a state model in terms of $v_c$ and $i_l$.
Note that, whatever the conditions allowing this are, the scope of this approach is in any case
restricted by the initial assumptions on the form of the characteristics (namely,
the current- and voltage-control assumptions above, or any other analogous ones).

Our key point is that we can do the same  
in terms of (\ref{core}), but now
getting an equivalent scenario {\em without} any control assumptions on the characteristics.
%or, equivalently, of (\ref{core-diff})-(\ref{core-alg}).
Incidentally, it is not by chance that  (\ref{core}) and (\ref{red01}) have the same structure: we can get (\ref{red01}) as a particular case of the general model (\ref{core})
in light of the assumptions above 
%When a global explicit description exists for a given device or set of devices... [EXPLAIN BETTER] (e.g.\
%the control assumptions above are accommodated in (\ref{core}) just by setting 
just by setting $u_r=v_r$
(that is, $\zeta_r(u_r)=u_r$) and then $\psi_r(v_r)=\xi_r(v_r)$, etc. 
But, as indicated above, 
the difference between both approaches is that (\ref{core}) does not require any {\em a priori}
control assumptions on the characteristics.

Actually, in the homogeneous framework we can easily formulate the state-space problem
as follows: the goal is to express $u_r$ in terms of $u_c$ and $u_l$ from (\ref{core-alg}), 
so that the insertion of the resulting expressions %$u_r$ in these terms with
in (\ref{core-diff}) %would %essentially
yields the desired state-space reduction. Needless to say, once the trajectories  are computed in terms
of the homogeneous variables $u$, we get the corresponding values of the
 classical circuit variables simply via $\psi_c$, $\zeta_c$ in (\ref{homog-cap2}), etc., 
which can be understood to be output maps (in the terminology of control theory).

%as $v_c=\zeta_c(u_c)$, $i_r=\psi_r(u_r)$, etc.

In this setting, the state-space reduction problem actually involves 
three different aspects which we present in the sequel and tackle in later subsections. 
First, since
the trajectories of the circuit model (\ref{core}) (or, equivalently, of 
(\ref{core-diff})-(\ref{core-alg}))
are explicitly bound to lie on the set defined by (\ref{core-alg}),
it is important in practice to examine when these equations
define a smooth manifold. Borrowing the term from the differential-algebraic literature,
we will call the
set defined by (\ref{core-alg}) the {\em constraint set} and
denote it by ${\cal M}$. 

Second, as indicated above, the
most natural approach to address the state-space problem is to express the
variables $u_r$ in terms of $u_c$, $u_l$.
Because of the linearity of Kirchhoff laws, we will be able to assess the
conditions for this
independently of the constraint set requirement above, specifically
by examining a non-singularity condition on the 
matrix of partial derivatives of the equations in the
left-hand side of (\ref{core-alg}) with respect to the variables
$u_r$. This will be the key ingredient in the definition of the
{\em regular set} ${\cal R}$. 
%Points in the homogeneous subspace $\mathbb{H}_r$ which satisfy (resp.\
%do not satisfy) this condition are termed {\em regular} (resp.\
%{\em singular}). Together with regular points within $\mathbb{H}_c \times \mathbb{H}_l$,
%which are just those yielding non-zero leading
%coefficients in (\ref{core-diff}), we will be led to the so-called 
%{\em regular} and  {\em singular} sets on the whole homogeneous 
%space $\mathbb{H}$, to be denoted by
%% the set or regular (resp.\ singular) points in the
%%homogeneous space will be denoted by 
%${\cal R}$ and ${\cal S}$.

Finally, the intersection of the constraint set ${\cal M}$ 
and the regular set ${\cal R}$, which by construction is guaranteed 
to be a manifold, will be termed the {\em regular manifold} and
denoted by ${\cal M}_{\mathrm{reg}}$. The circuit equations yield a well-defined flow
(in the usual sense of dynamical systems theory: see e.g.\ \cite{amann}) on 
${\cal M}_{\mathrm{reg}}$. In our context, this set would correspond to the {\em index one}
set in the differential-algebraic literature (cf.\ \cite{LMTbook, wsbook}); be aware of the fact
that the index one context is due to the topological nondegeneracy hypothesis.

A  problem closely related to the latter one
involves %the structure of 
the intersection of the constraint set
and the singular set $\mathbb{H}-{\cal R}$, which defines the so-called {\em impasse set}.
Generically, at impasse points a pair of trajectories collapse with infinite
speed in (either forward or backward) finite time: 
cf.\ %the seminal references 
\cite{chu1, chu2}. 
%within the nonlinear circuit context. %and also \cite{rabrheintheo}. 
Other behaviors are however possible
and a taxonomy of dynamical phenomena is discussed in more general terms
in \cite{wsbook, sotzhJDE01}; note that in the
latter works
the term ``impasse'' %is understood as a synonym for ``singular'', going 
goes beyond the generic
collapsing behavior mentioned above. At impasse points there is no chance to describe 
the dynamics in terms of a state-space model formulated as
an explicit ODE, but a so-called quasilinear reduction captures the dynamics
(cf.\ \cite{rabrheintheo, wsbook} and the example in subsection \ref{subsec-manifold}
below).

%We address these problems in the following subsections.
%IF I FINALLY INCLUDE SING CROSS PHENOMENA: A closely related problem
%is the structure of the intersection of the constraint set
%and the singular set, which defines the so-called {\em impasse set}.
%The circuit equations yield a well-defined flow
%(in the usual sense of dynamical
%systems theory: see e.g.\ \cite{amann}) on 
%${\cal M}_{\mathrm{reg}}$, even if there may also be
%smooth trajectories defined through certain points of the 
%impasse set.

Note finally that all %Worth noting is the fact that 
the sets defined above lie on the homogeneous space $\mathbb{H}$. Via 
the maps (\ref{homog-res2}), (\ref{homog-cap2}) and (\ref{homog-ind2}) these
sets are easily recast in terms of the classical circuit variables.

\subsection{The constraint set, the regular set and the regular manifold}
\label{subsec-reg}

%Roswitha: ``constraint set''

%Maybe ``feasible set'' but too creative for this context

%Myself: index one... index two... ``solution manifold'' (but impasse
%set: there may be solutions through the singularity). 

As indicated above, the subset ${\cal M}$ of $\h = \h_c \times \h_l \times \h_r$ defined by 
(\ref{core-alg}) is called the {\em constraint set}. 
In general, this set is defined
by $m_r=m-(m_c+m_l)$ equations on the $m=m_c + m_l + m_r$ variables $u$.
Note
that in degenerate cases this may be an  empty set (think e.g.\ of a circuit
with two diodes in series which are oriented in opposite directions).
When this is not the case, the state-space problem (bound to topologically
nondegenerate contexts) %hypothesis)
may now be generally stated as the formulation of conditions on (\ref{core-alg})
under which the variables $u_r$ can be expressed (at least locally) 
in terms of $u_c$, $u_l$; this locally makes ${\cal M}$ a manifold
which can be parametrized
using these homogeneous reactive variables. This will make 
it possible to recast \eqref{core-diff} as a (quasilinear) differential system on $u_c$, $u_l$,
providing an explicit state-space
model for the dynamics on the subset of $\h_c \times \h_l$
where the leading coefficients of \eqref{core-diff} do not vanish.
%It is worth indicating, however, that t
There are however 
other contexts in which ${\cal M}$ may be guaranteed
to be a manifold: %even if the matrix (\ref{matrix1}) is singular.
%we distinguish these cases since
%in this setting there is no chance to parametrize
%${\cal M}$ in terms of $u_c$, $u_l$. 
%Even if this is mostly left for future study, a 
%In this setting a quasilinear description of the dynamics
%is still possible, and a digression in this
%regard will be presented in 
cf.\ subsection \ref{subsec-manifold} in this regard.

%As indicated above, in practice we are interested in cases in which
%the constraint set ${\cal M}$ defined above %not only is an 
%$(m_c+m_l)$-manifold which %but, additionally, it 
%can be parametrized, at least locally, in terms of the reactive homogeneous variables
%$u_c$, $u_l$. 
%Certainly, t
The natural way to describe locally ${\cal M}$
in terms of the reactive homogeneous variables
$u_c$, $u_l$ involves characterizing the %set of 
points where
the matrix of derivatives of the equations in the left-hand side of (\ref{core-alg}) w.r.t.
the variables $u_r$, that is,
%the points where 
\begin{eqnarray} \label{matrix1}
\begin{pmatrix} A_c^{\perp} A_r\psi_r'(u_r) \vspace{1mm}\\
  B_l^{\perp} B_r\zeta_r'(u_r)
\end{pmatrix},
\end{eqnarray}
is non-singular. Note that the structure of (\ref{core-alg})
(or, in essence, the linearity of Kirchhoff laws) 
makes this matrix of partial derivatives dependent only on $u_r$ and
not on $u_c$, $u_l$. Together with the fact that the coefficients
of $u_c'$ and $u_l'$ on (\ref{core-diff}) depend only
on $u_c$ and $u_l$, respectively, this will yield
a Cartesian product structure  on the regular set defined below.

%\vv

\begin{defin} \label{defin-reg}
We define the {\em regular set} ${\cal R} \subseteq \h=\h_c \times \h_l \times \h_r$ 
of the homogeneous model (\ref{core}) as the Cartesian product
${\cal R}_c \times {\cal R}_l \times {\cal R}_r$, where

\begin{itemize}
\item 
${\cal R}_c$ and ${\cal R}_l$ are the
sets of values of $u_c \in \h_c$ and $u_l\in \h_l$ 
where all the components of %coefficients 
$\psi_c'(u_c)$
and $\zeta_l'(u_l)$ 
%in (\ref{core-diff}) 
are non-null; and
\item 
${\cal R}_r$ is the set of values of $u_r \in \h_r$ where the
matrix (\ref{matrix1}) is non-singular.
\end{itemize}
%%  \begin{eqnarray*}
%%{\cal R}_c & = & \{u_c \in {\mathbb{H}_c} \ / \ p_{c_i} (u_{c_i}) \neq 0 \text%{ for all } i \in E_c\} \\
%%{\cal R}_l & = & \{u_l \in {\mathbb{H}_l} \ / \ q_{l_i} (u_{l_i}) \neq 0 \text%{ for all } i \in E_l\} \\  
%%{\cal R}_r & = & \{u_r \in {\mathbb{H}_r} \ / \ K(u_r) \neq 0\}. %%NOT YET
%%  \end{eqnarray*}

%\vs

\noindent The set $\h - {\cal R}$ is called the {\em singular set}.
\end{defin}

\vs

\noindent Mind the terminological abuse: $\psi_c'(u_c)$
and $\zeta_l'(u_l)$, as matrices of partial derivatives, are diagonal because
of the absence of coupling effects, and by their components we 
mean the diagonal entries of such matrices, namely, the
derivatives $\psi_{c_i}'$
and $\zeta_{l_j}'$ (depending on $u_{c_i}$ and $u_{l_j}$, respectively), $i$ and $j$
indexing the sets of capacitors and inductors, respectively.

The only factor in the regular set which is not 
%trivially defined 
explicitly described
%characterized 
in Definition \ref{defin-reg}
is the (say) ``resistive'' regular set ${\cal R}_r$. More precisely, the problem
here is to characterize this set in structural terms, that is, in terms
of the topology of the circuit graph and the electrical features
of the devices.
In Theorem \ref{th-reg} below, these circuit-theoretic terms involve
the structure of the circuit spanning trees: 
specifically, we make use of the notion of a {\em proper tree},  
%(a concept which can be traced back to \cite{bashkow57}), 
which
is a spanning tree including all capacitors and no inductor.  The existence of 
at least one proper tree is a 
well-known consequence of the topological nondegeneracy hypothesis.
%(a property which was explicitly proved already in \cite{brown63}). 
The set of proper
trees of a given circuit will be denoted  by ${\cal T}_p$,
whereas ${\cal T}$ denotes the family of all spanning trees.
In Theorem \ref{th-reg} we denote by $E_r$ the index set of resistive branches: this way, 
$T \cap E_r$ and $\overline{T} \cap E_r$ stand, respectively, for
the index sets of the resistive branches within a given tree $T$ and of
those in the corresponding cotree. Additionally, we assume w.l.o.g.\ that resistive
branches are the first $m_r$ ones, so that 
$p_{r_i}$ and $q_{r_i}$ denote
the derivatives of the $i$-th component
of $\psi_r$ and $\zeta_r$ in (\ref{homog-res2}); note that both derivatives depend only on $u_{r_i}$.
%********* TO MAKE THE NOTATION CONSISTENT $E_r$ SHOULD BE THE FIRST $m_r$ BRANCHES
%The proof of this statement is essentially based on the  results presented 

%NOT HERE: We emphasize the fact that the generality of this result is a consequence of the homogeneous formalism.
%The proof makes use of an auxiliary result proved in
%along the lines introduced in 
%\cite{homoglin}.

\vs

\begin{theor} \label{th-reg}
  The
set ${\cal R}_r \subseteq \h_r$ is explicitly characterized
by the non-vanishing of 
the function
%polynomial %determinant $K(u_r)$ of (\ref{matrix1}) reads as %equals the sum
  \begin{eqnarray} \label{kir-nonl}
    K(u_r) = \sum_{T \in {\cal T}_p}\left(\hspace{0.5mm} \prod_{i \in T \cap E_r}p_{r_i}(u_{r_i}) \prod_{j \in \overline{T} \cap E_r}q_{r_j}(u_{r_j})\right).
    \end{eqnarray}
  %The regular set is $\mathbb{H}_c \times \mathbb{H}_l \times {\cal R}_r$,
  %where ${\cal R}_r$ is the set where 
  %the multihomogeneous Kirchhoff polynomial (**) does not vanish.
\end{theor}

\noindent The proof will be based on the following auxiliary 
result (cf.\  \cite[Theorem 1]{homoglin}), 
which can be understood as a projectively-weighted
version of the matrix-tree theorem.

\vs

\begin{lema} \label{lema-lin}
Assume that $A$ and $B$ are, respectively, % two given 
a reduced cut matrix (or an incidence matrix) 
and a cycle matrix of a connected digraph.
Let $P$, $Q$ be arbitrary diagonal matrices, with $p=(p_1, \ldots, p_m)$ and
$q=(q_1, \ldots, q_m)$ the vectors of diagonal entries of $P$ and $Q$.
Then
\begin{equation} \label{kir-hom}
\det 
\begin{pmatrix} AP \\ BQ \end{pmatrix}
=k_{AB}\sum_{T \in {\cal T}} 
\left( \prod_{i \in T} p_i \prod_{j \in \overline{T}} q_j\right), 
\end{equation}
for a certain non-zero
constant $k_{AB}$.
\end{lema}

\vs

\noindent Disregarding the $k_{AB}$ factor, 
the function in the right-hand side of (\ref{kir-hom})
is the so-called multihomogeneous Kirchhoff (or tree-enumerator)
polynomial of a connected graph, 
to be denoted by $\tilde{K}(p,q)$, in which
%which is...
%**homoglin the multihomogeneous Kirchhoff polynomial of a connected
%(di)graph (see \cite{chaiken, smith72} and references therein) is defined as
%\begin{equation} \label{kir-hombis}
%K(p,q)=\sum_{T \in {\cal T}} 
%\left( \prod_{j \in T} p_j \prod_{k \in \overline{T}} q_k\right),
%\end{equation}
%that is,
every spanning tree $T$ sets up a monomial
which includes $p_i$ (resp.\ $q_i$) as a factor if the $i$-th branch
belongs to $T$ (resp.\ %belongs 
to %the cotree 
$\overline{T}$) \cite{chaiken, homoglin} %; regarding this concept,
(the example discussed below can be of help for the reader at this point).

\vv

\noindent {\bf Proof of Theorem \ref{th-reg}.}
With the splitting $A=(A_c \ A_l \ A_r)$, $B=(B_c \ B_l \ B_r)$,
and by setting $P=$ block-diag$(I_{m_c}, \ 0_{m_l}, \ \psi_r'(u_r))$,
$Q=$ block-diag$(0_{m_c}, \ I_{m_l}, \ \zeta_r'(u_r))$, the matrix in the
left-hand side of (\ref{kir-hom}) reads as
\begin{equation} \label{threg1}
\begin{pmatrix} AP \\ BQ \end{pmatrix}=
\begin{pmatrix} A_c & 0 & A_r\psi_r'(u_r) \\ 0 & B_l & B_r\zeta_r'(u_r)\end{pmatrix}.
\end{equation}
By Lemma \ref{lema-lin}, the determinant of this matrix is defined by the polynomial
in the right-hand side of \eqref{kir-hom}. 
Because of the definition of the $P$ matrix,
all values of $p$ corresponding to inductors do vanish, 
whereas for capacitors we have $p_{c_i}=1$;
dually, values of $q$ which correspond to capacitors are null, and for inductors we have $q_{l_i}=1$.
This means that  any inductor belonging to a tree annihilates
the corresponding term in the Kirchhoff polynomial, because of the vanishing 
of $p_{l_i}$; 
analogously, any capacitor in a cotree renders the term for that tree null, 
since $q_{c_i}=0$. Therefore, the only (possibly) non-null terms in the 
polynomial
must correspond to proper trees, namely, trees including all capacitors and no inductor. Note, 
additionally, that within these trees we have $p_{c_i}=1$ and $q_{l_i}=1$, so that only the
resistive terms actually contribute a (possibly) nontrivial factor within 
each monomial. Altogether, this means that the determinant
of (\ref{threg1}) equals $k_{AB}K(u_r)$, with the latter function
defined in (\ref{kir-nonl}).
%identity
%  \begin{eqnarray}
%    \det \begin{pmatrix} AP \\ BQ \end{pmatrix} = k_{AB} \sum_{T \in {\cal T}_p}\left(\hspace{0.5mm} \prod_{j \in T \cap E_r}p_{r_j}(u_{r_j}) \prod_{k \in \overline{T} \cap E_r}q_{r_k}(u_{r_k})\right)
%    \end{eqnarray}
%holds for the matrix in the left-hand side of (\ref{threg1}).

%try mathbf perp

It remains to show that, except for another non-null factor, the determinant
of %the right-hand side of 
(\ref{threg1}) %amounts to 
equals that of (\ref{matrix1}).
To check this we premultiply 
%use the %aforementioned 
%fact that, in light of the topological nondegeneracy hypothesis and the
%definition of $A_c^{\perp}$ and $B_l^{\perp}$, the matrices 
%\begin{eqnarray} \label{ACBLbis}
%\tilde{A}=\begin{pmatrix}A_c^{\tra} \vspace{0.5mm} \\ A_c^{\perp} \end{pmatrix} 
%\text{ and }
%\tilde{B}= \begin{pmatrix}B_l^{\perp} \vspace{0.5mm} \\ B_l^{\tra} \end{pmatrix}
%\end{eqnarray}
%%[or maybe the ones from (\ref{ACBL}), change in that case], 
%are non-singular.
%Now, premultiply 
the right-hand side of (\ref{threg1})
%by the matrix block-diag($\tilde{A}$, $\tilde{B}$),
by the matrix block-diag($A_0$, $B_0$) (cf.\ (\ref{ACBL})),
%\begin{equation}\label{matrixAux}
%\begin{pmatrix}A_c^{\tra} \vspace{0.5mm} & 0 \\ A_c^{\perp} & 0 \\ 
%0 & B_l^{\perp} \vspace{0.5mm} \\ 0 & B_l^{\tra} \end{pmatrix},
%\end{equation}
which is %itself 
non-singular by construction, to get
%First to second version change A_c^{\tra} by A_c^- and the same for B
\ifthenelse{\boolean{ieee}}{
%$$\begin{pmatrix}A_c^- \vspace{0.5mm} & 0 \\ A_c^{\perp} & 0 \\ 
%0 & B_l^{\perp} \vspace{0.5mm} \\ 0 & B_l^- \end{pmatrix}
%\begin{pmatrix} A_c & 0 & A_r\psi_r'(u_r) \\ 0 & B_l & B_r\zeta_r'(u_r)\end{pmatrix}=\hspace{20mm}$$
$$ %\ \hspace{10mm}=
\begin{pmatrix} I_{m_c} & 0 & A_c^-A_r\psi_r'(u_r) \\ 
0 & 0 & A_c^{\perp}A_r\psi_r'(u_r) \\ 
0 & 0 & B_l^{\perp}B_r\zeta_r'(u_r)\\
0 & I_{m_l} & B_l^- B_r\zeta_r'(u_r)
\end{pmatrix},
$$
}{
$$
\begin{pmatrix} I_{m_c} & 0 & A_c^-A_r\psi_r'(u_r) \\ 
0 & 0 & A_c^{\perp}A_r\psi_r'(u_r) \\ 
0 & 0 & B_l^{\perp}B_r\zeta_r'(u_r)\\
0 & I_{m_l} & B_l^- B_r\zeta_r'(u_r)
\end{pmatrix},
$$
}
whose determinant equals, maybe up to a sign, that of (\ref{matrix1}).
%$$\pm %\det \left( A_c^- A_c \right)\det \left( B_l^- B_l \right) 
%\det
%\begin{pmatrix} A_c^{\perp} A_r\psi_r'(u_r) \vspace{1mm}\\
%  B_l^{\perp} B_r\zeta_r'(u_r)
%\end{pmatrix}.
%$$
%We finally make use of 
%Using the fact that $\det \left( A_c^{\tra} A_c \right)\det \left( B_l^{\tra} B_l \right) \neq 0$,
%owing to the absence of C-loops and L-cutsets
%(which makes the columns of $A_c$ and $B_l$  linearly independent),
It then follows %from the latter expression 
that (\ref{matrix1}) and (\ref{threg1}) actually have
(possibly up to a non-null factor) the same determinant, as claimed.
%the function defined in (\ref{kir-nonl}) 
%and the claim is proved.
\hfill $\Box$

%(maybe define $K_c$ as the product of the $p_c$'s etc.

\vs

%MAYBE AN EXAMPLE OR COUPLE OF EXAMPLES... 

%- Chua's circuit: both resistors in the cotree, impasse if $R=0$ there... 

%- maybe a BRIDGE

%- Zhou's paper in JCSC 24:9 2015 (fractional order, have PDF). maybe for THEVENIN. Mention:

%- (check Google) equivalent circuit graphene

\ifthenelse{\boolean{ieee}}{{\it Example. Murali-Lakshmanan-Chua circuits.}}{\paragraph{Example. Murali-Lakshmanan-Chua circuits.}} A key role in the result above is played
by the polynomial in the right-hand side of (\ref{kir-hom}) and its nonlinear counterpart
(\ref{kir-nonl}).
We illustrate 
%the scope of Theorem \ref{th-reg} and 
the form that these functions
take in practice by means on an example defined by two resistively-coupled 
Murali-Lakshmanan-Chua (MLC) circuits, depicted in Fig.\ 2.
MLC circuits were introduced in \cite{murali94}, and arrays of these
circuits are considered for different purposes 
e.g.\ in \cite{laks2013, muruga99}; see also \cite{ishaq2017}. We use 
one of the circuits of the MLC family defined in \cite{laks2013};
to focus on the contribution of resistors we set $C=L=1$ and 
annihilate the voltage in voltage sources within the original 
circuit as defined in that paper. 

\ifthenelse{\boolean{ieee}}{\begin{figure}[h]
\vspace{4mm}
\centering
%\mbox{} \hspace{-6mm} 
%\parbox{2.0in}{
%\ \mbox{} \hspace{-25mm} %\vspace{5.5mm}
\epsfig{figure=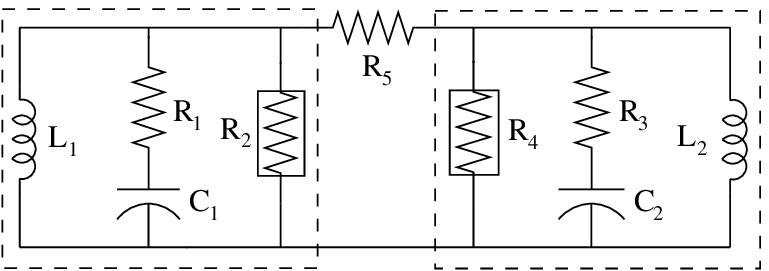, width=0.45\textwidth}
%}%, angle=270}
\vspace{1mm}
\caption{Coupled Murali-Lakshmanan-Chua circuits.\hspace{1mm}}
\label{fig-MLC}
%\vspace{1mm}
\end{figure}}
{\begin{figure}[h]
\vspace{4mm}
\centering
%\mbox{} \hspace{-6mm} 
%\parbox{2.0in}{
%\ \mbox{} \hspace{-25mm} %\vspace{5.5mm}
\epsfig{figure=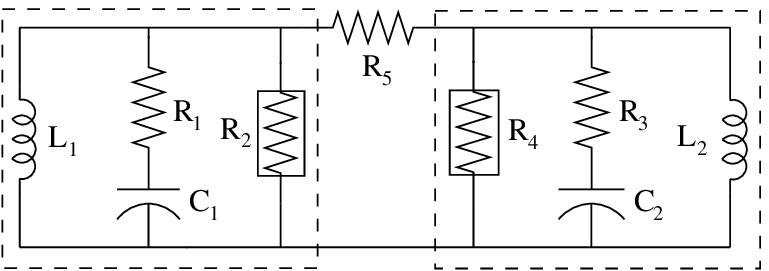, width=0.60\textwidth}
%}%, angle=270}
\vspace{1mm}
\caption{Coupled Murali-Lakshmanan-Chua circuits.\hspace{1mm}}
\label{fig-MLC}
%\vspace{1mm}
\end{figure}}

\ifthenelse{\boolean{ieee}}{
\begin{figure}[h]
\vspace{2mm}
%\centering
\mbox{} \hspace{-1mm} 
\parbox{1.8in}{
\ \mbox{} \hspace{1mm} %\vspace{5.5mm}
\epsfig{figure=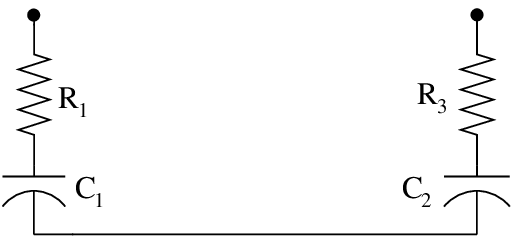, width=0.24\textwidth}}%, angle=270}
%\ \mbox{} 
%\hspace{1mm} %\vspace{5.5mm}
\parbox{1.8in}{
\epsfig{figure=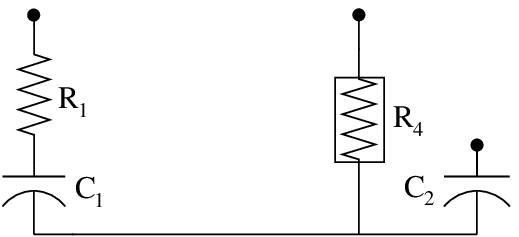, width=0.24\textwidth}\hspace{-2mm}}%, angle=270}
%\ \mbox{} 
\vspace{6mm}\\
\mbox{} \hspace{-1mm} 
\parbox{1.8in}{
\ \mbox{} \hspace{1mm} %\vspace{5.5mm}
\epsfig{figure=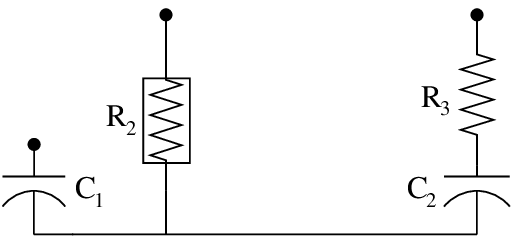, width=0.24\textwidth}}%, angle=270}
%\hspace{1mm} 
\parbox{1.8in}{
\epsfig{figure=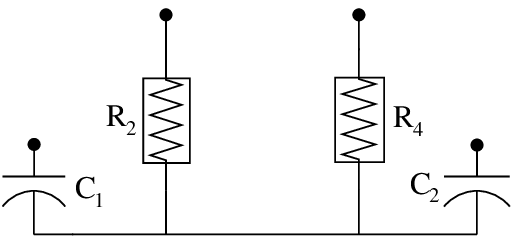, width=0.24\textwidth}}%, angle=270}
%\ \mbox{} 
\vspace{6mm}\\
\mbox{} \hspace{-1mm} 
\parbox{1.8in}{
\ \mbox{} \hspace{1mm} %\vspace{5.5mm}
\epsfig{figure=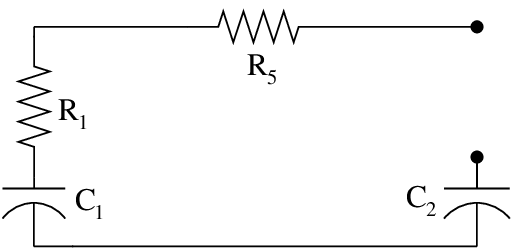, width=0.24\textwidth}}%, angle=270}
%\ \mbox{} 
%\hspace{1mm} 
\parbox{1.8in}{
%\vspace{5.5mm}
\epsfig{figure=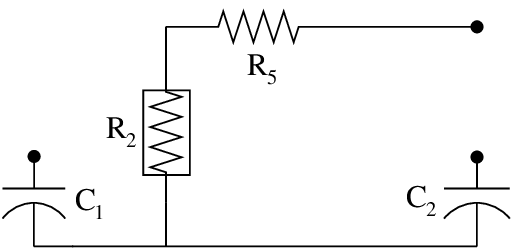, width=0.24\textwidth}}%, angle=270}
\vspace{6mm}\\
\mbox{} \hspace{-1mm} 
\parbox{1.8in}{
\ \mbox{} \hspace{1mm} %\vspace{5.5mm}
\epsfig{figure=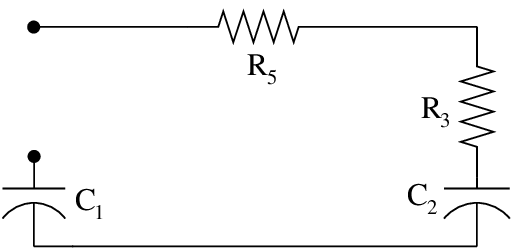, width=0.24\textwidth}}%, angle=270}
%\hspace{1mm} 
\parbox{1.8in}{
\epsfig{figure=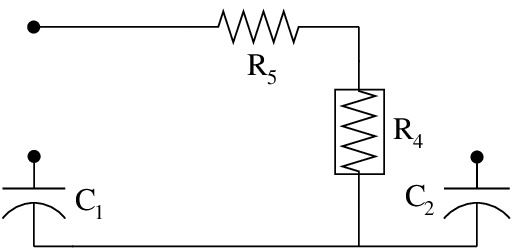, width=0.24\textwidth}}%, angle=270}
\vspace{3mm}
\caption{Proper trees.}
\label{fig-proper}
\end{figure}
}{
\begin{figure}[h]
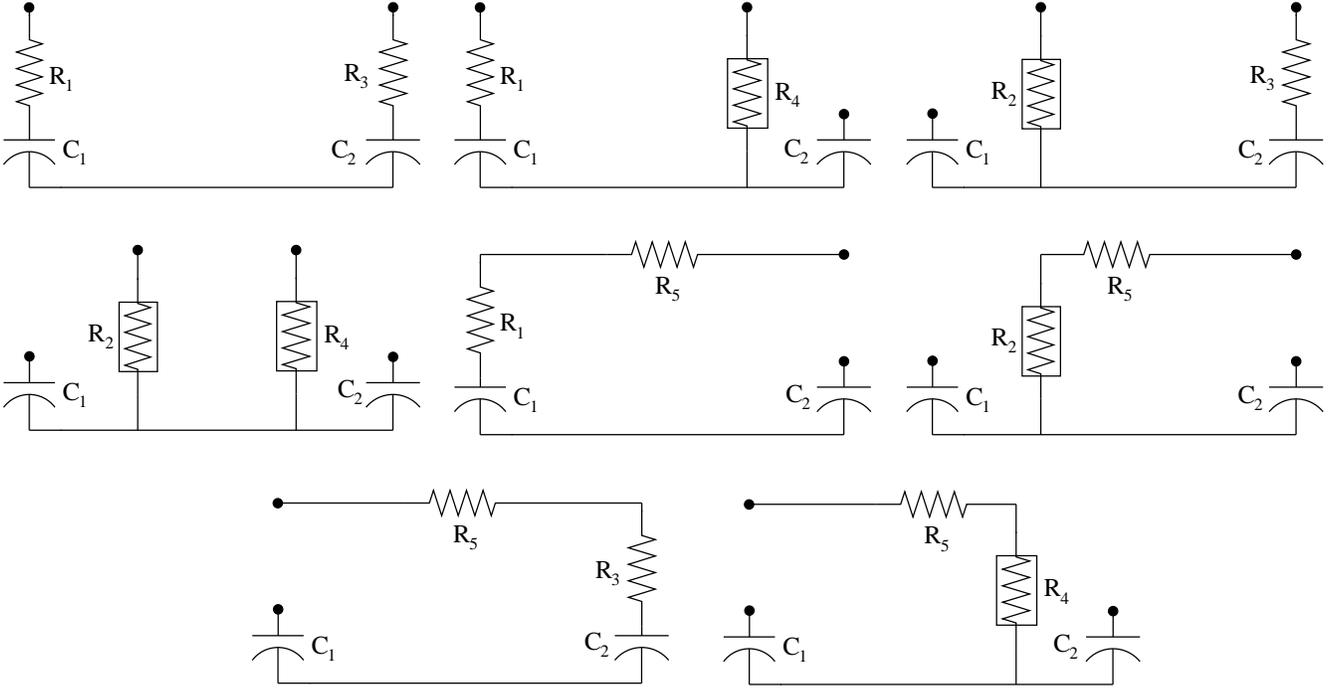

\vspace{4mm}
\centering
%\mbox{} \hspace{-6mm} 
\parbox{2.0in}{
\ \mbox{} \hspace{-10mm} %\vspace{5.5mm}
\epsfig{figure=tree13, width=0.34\textwidth}}%, angle=270}
%\ \mbox{} 
\hspace{2mm} %\vspace{5.5mm}
\parbox{2.0in}{
\epsfig{figure=tree14, width=0.34\textwidth}}%, angle=270}
%\ \mbox{} 
\hspace{8mm} \parbox{2.0in}{
%\vspace{5.5mm}
\epsfig{figure=tree23, width=0.34\textwidth}}%, angle=270}
\vspace{6mm}\\
\parbox{2.0in}{
\ \mbox{} \hspace{-10mm} %\vspace{5.5mm}
\epsfig{figure=tree24, width=0.34\textwidth}}%, angle=270}
%\ \mbox{} 
\hspace{2mm} \parbox{2.0in}{
%\vspace{5.5mm}
\epsfig{figure=tree15, width=0.34\textwidth}}%, angle=270}
%\ \mbox{} 
\hspace{8mm} \parbox{2.0in}{
%\vspace{5.5mm}
\epsfig{figure=tree25, width=0.34\textwidth}}%, angle=270}
\vspace{6mm}\\
\mbox{} \hspace{12mm} 
\parbox{2.0in}{
\ \mbox{} \hspace{-16mm} %\vspace{5.5mm}
\epsfig{figure=tree35, width=0.34\textwidth}}%, angle=270}
\parbox{2.0in}{
\ \mbox{} \hspace{-6mm} %\vspace{5.5mm}
\epsfig{figure=tree45, width=0.34\textwidth}}%, angle=270}
\vspace{1mm}
\caption{Proper trees.}
\label{fig-proper}
\end{figure}
}

From the set of proper trees (displayed in Fig.\ 3)
one can easily check that the 
multihomogeneous Kirchhoff polynomial reads for this circuit as
\ifthenelse{\boolean{ieee}}{\begin{eqnarray}
%\hspace{-3mm}
\nonumber p_1q_2p_3q_4q_5 + %\hspace{-0.5mm}+ \hspace{-0.5mm} 
p_1q_2q_3p_4q_5 + %\hspace{-0.5mm}+ \hspace{-0.5mm}  
q_1p_2p_3q_4q_5 + %\hspace{-0.5mm}+ \hspace{-0.5mm}  
q_1p_2q_3p_4q_5 + \hspace{1mm} \\ +%\hspace{-0.5mm}+ \hspace{-0.5mm}  
p_1q_2q_3q_4p_5 +  %\hspace{-0.5mm}+ \hspace{-0.5mm}  
q_1p_2q_3q_4p_5 + %\hspace{-0.5mm}+ \hspace{-0.5mm}  
q_1q_2p_3q_4p_5 + %\hspace{-0.5mm}+ \hspace{-0.5mm}  
q_1q_2q_3p_4p_5.\ 
\label{polex}
\end{eqnarray}}{\begin{eqnarray}
%\hspace{-3mm}
\nonumber  p_1q_2p_3q_4q_5 + %\hspace{-0.5mm}+ \hspace{-0.5mm} 
p_1q_2q_3p_4q_5 + %\hspace{-0.5mm}+ \hspace{-0.5mm}  
q_1p_2p_3q_4q_5 + %\hspace{-0.5mm}+ \hspace{-0.5mm}  
q_1p_2q_3p_4q_5 + %\hspace{-0.5mm}+ \hspace{-0.5mm}  
p_1q_2q_3q_4p_5 + \hspace{16.5mm} \\ + %\hspace{-0.5mm}+ \hspace{-0.5mm}  
q_1p_2q_3q_4p_5 + %\hspace{-0.5mm}+ \hspace{-0.5mm}  
q_1q_2p_3q_4p_5 + %\hspace{-0.5mm}+ \hspace{-0.5mm}  
q_1q_2q_3p_4p_5. 
\label{polex}
\end{eqnarray}}
The function (\ref{kir-nonl}), characterizing the set of regular points,
is just obtained by letting $p_i$ and $q_i$ 
above depend on the corresponding homogeneous variable $u_i$. 
We emphasize the fact that the non-vanishing of this
function of the homogeneous variables performs this characterization of
the regular set in full generality.
It is of
interest, however, to show how this general model 
takes simpler forms and provides additional
information in simplified settings which arise
from different assumptions on the circuit devices, as we do in the sequel.

Indeed, in each MLC circuit only one of the resistors
displays a nonlinear behavior (namely, those labelled with the subindices
2 and 4), whereas numbers 1 and 3, as well as the coupling resistor
5, are typically linear. If, moreover, we assume them to be defined
by a resistance parameter $r_i$, $i=1, 3, 5$ (this is equivalent
to saying that $p_1$, $p_3$ and $p_5$  %$p_2$ and $p_3$ 
do not vanish) we may divide 
%and then, by dividing
the polynomial above by $p_1p_3p_5$ to get a partially dehomogenized
form which characterizes the regular set of values for the remaining
homogeneous variables (namely, $u_2$ and $u_4$); note that $r_i =q_i/p_i$.
These are defined by the non-vanishing of the function
(we group some terms for notational simplicity):
\ifthenelse{\boolean{ieee}}{\begin{eqnarray*}
%\hspace{-3mm}
p_2(u_2)p_4(u_4) r_1r_3r_5 + %\hspace{-0.5mm}+ \hspace{-0.5mm}  
p_2(u_2)q_4(u_4)(r_1r_3 + %\hspace{-0.5mm}+\hspace{-0.5mm} 
r_1 r_5) 
+ %\hspace{-0.5mm}+ \hspace{-0.5mm}
\hspace{15mm} % \hspace{-0.5mm}+ 
\\ %\hspace{-8mm}  
+ q_2(u_2)p_4(u_4) (r_1r_3 + %\hspace{-0.5mm}+\hspace{-0.5mm} 
r_3 r_5)
+ q_2(u_2)q_4(u_4) (r_1 + % \hspace{-0.5mm}+ \hspace{-0.5mm}  
r_3 + %\hspace{-0.5mm}+ \hspace{-0.5mm}  
r_5). \hspace{3mm}
\end{eqnarray*}}{\begin{eqnarray*}
\hspace{-3mm}
p_2(u_2)p_4(u_4) r_1r_3r_5 + %\hspace{-0.5mm}+ \hspace{-0.5mm}  
p_2(u_2)q_4(u_4)(r_1r_3 + %\hspace{-0.5mm}+\hspace{-0.5mm} 
r_1 r_5) 
+ %\hspace{-0.5mm}+ \hspace{-0.5mm}
q_2(u_2)p_4(u_4) (r_1r_3 + %\hspace{-0.5mm}+\hspace{-0.5mm} 
r_3 r_5)
+ \hspace{15mm} % \hspace{-0.5mm}+ 
\\ %\hspace{-0.5mm}  
+ \ q_2(u_2)q_4(u_4) (r_1 + % \hspace{-0.5mm}+ \hspace{-0.5mm}  
r_3 + %\hspace{-0.5mm}+ \hspace{-0.5mm}  
r_5).
\end{eqnarray*}}
%Note that i
In the latter formula we retain an homogeneous expression
for both nonlinear resistors. Still by way of example,
assume now that resistor no.\ 4 is known to admit
a global voltage-controlled expression: $u_4$ then amounts to the voltage
variable $v_4$ and the expression above may be divided by $q_4$
to get a description of this device in terms of the incremental conductance $g_4(v_4)$.
For resistor no.\ 2 we retain, by contrast,
the homogeneous form, for instance to keep the chance to model 
both an open-circuit and a short-circuit for this resistor (this may be relevant 
in fault diagnosis applications, since
both situations may arise in faulty circuits).
%(that is, an unexpected short-circuit). 
Under these hypotheses, the
function characterizing the regular set is %would be
\ifthenelse{\boolean{ieee}}{\begin{eqnarray}
 p_2(u_2)g_4(v_4) r_1r_3r_5 
\hspace{0.0mm} + \hspace{0.0mm}  
p_2(u_2)(r_1r_3 \hspace{0.0mm}+\hspace{0.0mm} 
r_1 r_5) 
\hspace{0.0mm} + \hspace{15mm} % \hspace{0.0mm}+ 
\nonumber \\ \ + \hspace{0.0mm}
q_2(u_2)g_4(v_4) (r_1r_3 \hspace{0.0mm}+\hspace{0.0mm} r_3 r_5)
\hspace{0.0mm}  %\hspace{0.0mm}  
+ \hspace{0.0mm}  
q_2(u_2)(r_1 
\hspace{0.0mm} + \hspace{0.0mm} r_3 
\hspace{0.0mm}+ \hspace{0.0mm} r_5). %\ \ \ 
\ \label{exaux}
\end{eqnarray}}
{\begin{eqnarray}
 p_2(u_2)g_4(v_4) r_1r_3r_5 
\hspace{-0.5mm}+ \hspace{-0.5mm}  
p_2(u_2)(r_1r_3 \hspace{-0.5mm}+\hspace{-0.5mm} 
r_1 r_5) 
\hspace{-0.5mm} + \hspace{-0.5mm}
q_2(u_2)g_4(v_4) (r_1r_3 \hspace{-0.5mm}+\hspace{-0.5mm} r_3 r_5)
\hspace{-0.5mm} + \hspace{-0.5mm}  
q_2(u_2)(r_1 
\hspace{-0.5mm} + \hspace{-0.5mm} r_3 
\hspace{-0.5mm}+ \hspace{-0.5mm} r_5). %\ \ \ 
\label{exaux}
\end{eqnarray}}
Finally, a fault due to a short-circuit  in the
second resistor would be modeled here by $q_2=0$ (implying $p_2 \neq 0$). 
In this particular
context, the set of singular values for the remaining variable $v_4$
would simply be obtained by annihilating (\ref{exaux}), and
are given by 
%\begin{eqnarray}
%\hspace{53.5mm} 
$g_4(v_4) = -(r_1r_3 + r_1 r_5)/(r_1 r_3 r_5).$ 
%\hspace{53.5mm} 
Needless to say, other conclusions could be analogously drawn in other
working scenarios from the general form of the
multihomogeneous Kirchhoff polynomial (\ref{polex}).
\hfill $\Box$
% \label{exaux2}
%\end{eqnarray}

\vv

\noindent We finish this subsection with 
%by writing locally $u_r$ in terms of $u_c$, $u_l$ on the regular set, we get 
the following result, which essentially
says that a flow is well-defined
on the intersection ${\cal M} \cap {\cal R}$.
It is an immediate consequence of the non-singularity of (\ref{matrix1})
and the implicit function theorem, which yields a
local description of ${\cal M}$  in the form $u_r=\eta_r(u_c,u_l)$
near regular points. An elementary example of a state-space model of the form
(\ref{state}) below can be found in (\ref{vdp2}); note that 
the homogeneous variable $u_c$ amounts there to $v_c$ because
of the working assumptions in that example.

\vs

\begin{theor} 
If non-empty, 
the intersection of the constraint 
set ${\cal M}$ defined by (\ref{core-alg}) and
the regular set ${\cal R}$ in Definition \ref{defin-reg} %defined above
is an $(m_c+m_l)$-dimensional manifold. It is filled by solutions
of the circuit equations (\ref{core})
(or, equivalently, of (\ref{core-diff})-(\ref{core-alg})), which
are locally defined by the solutions of an explicit state-space model of the form
\ifthenelse{\boolean{ieee}}{\begin{subequations} \label{state}
\begin{eqnarray}
\mbox{} \hspace{-7mm}u_c' \hspace{-2mm}& = & \hspace{-2mm}-(\psi_c'(u_c))^{-1}A_c^-\left( A_l \psi_l(u_l) + A_r\psi_r(\eta_r(u_c,u_l))\right)\\
\mbox{} \hspace{-7mm}u_l' \hspace{-2mm}& = & \hspace{-2mm}-(\zeta_l'(u_l))^{-1}B_l^- \left(B_c\zeta_c(u_c) + B_r\zeta_r(\eta_r(u_c,u_l))\right).\
\end{eqnarray}
\end{subequations}}{\begin{subequations} \label{state}
\begin{eqnarray}
u_c' & = & -(\psi_c'(u_c))^{-1}A_c^-\left( A_l \psi_l(u_l) + A_r\psi_r(\eta_r(u_c,u_l))\right)\\
u_l' & = & -(\zeta_l'(u_l))^{-1}B_l^- \left(B_c\zeta_c(u_c) + B_r\zeta_r(\eta_r(u_c,u_l))\right).
\end{eqnarray}
\end{subequations}}
\end{theor}

\vs

\noindent 
%We will call t
The intersection ${\cal M} \cap {\cal R}$ yields
the regular manifold
%denoted  by 
${\cal M}_{\mathrm{reg}}$. This corresponds to the {\em index one}
set in the differential-algebraic literature (cf.\ \cite{LMTbook, wsbook, carentop}):
note that %the focus is restricted to an 
the index one context arises from %because of 
the 
assumed topological nondegeneracy.
%In the DAE literature this is known as the {\em index one} set.
The set ${\cal M}-{\cal M}_{\mathrm{reg}}$ is
%will be called 
the  impasse set. % (MAYBE LEAVE THIS DEFINITION FOR LOCALLY NONLINEAR PROBLEMS)

%A related problem: when there is a global state space equation 
%{\bf (think of the form of the state space above; i.f.t. is local but we are always
%solving in terms of $u_c$ and $u_l$)}. Think of VdPol example

%- when the regular set is open dense (in the underlying space); locally nonlinear thing. Below. Ok.

\subsection{The regular set is dense in locally nonlinear problems}
%The linear/nonlinear thing. Impasse set of nonlinear problems}

%\subsubsection{Regular and singular points in linear/nonlinear circuits}

%PREVIOUS: Motivate with example VdPol (CAREFUL THIS WAS FOR THE SERIES CONFIGURATION);
% to get more easily to the point
%let as assume that the inductor is a linear current-controlled one ($u_l$ amounts to $i_l$ or, in other words, 
%$\psi_l(i_l)=i_l$). 

In this subsection we elaborate on the structure of the impasse set
defined above. In order to motivate the discussion, 
let us go back to the partially homogeneous form of the Van der Pol system 
(with a linear capacitor) defined by (\ref{vdp1}).
The regular set in this case is defined by the conditions $\zeta_l'(u_l) \neq 0$
and $\zeta_r'(u_r) \neq 0$: we note in passing that this parallel configuration 
has a unique proper tree, just 
defined by the capacitor; the resistor is therefore in the cotree and hence the 
latter condition on $\zeta_r'(u_r)=q_r(u_r)$.
%(the latter is $q(u_r) \neq 0$, with $q$ from within the
%unique proper cotree)
%[Constraint: still a manifold if $\psi_l'(u_l) \neq 0$ - possibly simpler setting by assuming linear L]
Now, for a generic set of functions $\zeta_l$ and $\zeta_r$
(think e.g.\ of Morse functions,
for which %all critical points are non-degenerate; this means that 
the condition $\zeta'(u)=0$ implies
$\zeta''(u) \neq 0$, making all critical points
isolated), the singular set is simply defined by a set of hyperplanes
of the form $u_l=u_l^*$ and $u_r=u_r^*$, where
$u_l^*$ and $u_r^*$ 
denote critical points of $\zeta_l$ and $\zeta_r$, respectively.
The impasse set
is in this case a hypersurface of the constraint set ${\cal M}$ 
defined by (\ref{vdp1c}).

%This 
%On the linear/nonlinear distinction in circuit theory}
%By contrast, t
The nature of the singular
set is radically different if the inductor and the
resistor in (\ref{vdp1}) are also
 assumed to be %a
linear. Indeed, suppose
both to be linear and current-controlled,
so that 
%, that the inductor
%is a current-controlled linear one (so that $\psi_l$ is the identity,
%that is, 
$u_l$ and $u_r$ amount to the currents $i_l$ and $i_r$, with
 $\zeta_l(i_l)=Li_l$ and $\zeta_r(i_r)=Ri_r$.
%and, analogously, the resistor is a linear
%voltage-controlled one, FOR THE SERIES...
%yielding $u_r=v_r$ (or $\zeta_r = \mathrm{id}$) and $i_r=\psi_r(v_r)=Gv_r$. 
For further simplicity, assume $C$ and $L$ not to vanish.
In this setting, the assumption
$R \neq 0$ makes all points regular, whereas when $R=0$ all 
points would be singular according
to Definition \ref{defin-reg}. 
%the definition given in subsection \ref{subsec-reg}. 
%(in this specific example we would actually be led to an index
%two DAE, but this is not in the scope of our present discussion).
%(in practice, in the latter setting the problem wouldn't be labelled as a singular one
%but as an index-two configuration)
In particular, there is no hypersurface of singular points in the
whole homogeneous space $\h$
or of impasse points in the constraint set ${\cal M}$
(which in this case is simply a hyperplane, namely the one
defined by the linear relation $v_c=Ri_r$, here expressed
in terms of classical circuit variables because $u_c=v_c$ and $u_r=i_r$).

It is well known in circuit theory
that linear problems do not exhibit impasse phenomena; that is, the behavior
described above, with 
all points having the same (regular or singular) nature, is always found in linear problems. 
Notice that
singular cases typically yield higher index DAE models.
This is a rather obvious consequence of the fact that the eventual singularity of 
the matrix (\ref{matrix1}) does
not depend on $u_r$ in linear cases, together with the
remark that the leading
coefficients of %the time-derivatives in 
(\ref{core-diff}) would be constant in a linear setting.
But we are now in a position
to give much more precise information about this:
%description of the extent to which this idea applies (certainly,
%not only for the Van der Pol example but in general); 
generically or, more specifically, for the locally nonlinear
functions defined below, the 
regular set is 
%[singular set is the complement  of]
an open dense subset of the homogeneous space, as it was the case
for the example (\ref{vdp1}) mentioned above.
%Note that it will not necessarily be true that the singular set (the complement of the
%regular set) is a hypersurface, that is, a codimension-one
%manifold, since it may well have self-intersections (think of a point of the form $(u_c, u_l^*, u_r^*)$
%in the example above). 
%It is interesting to note that this really makes a qualitative difference
%between linear circuits and (properly) nonlinear ones.

%\subsubsection{Locally nonlinear devices yield an open dense regular set}

From the theory of parametrized curves
we know that the curvature of a (regularly) parametrized curve 
$(\psi(u),  \zeta(u))$
at a given $u$ is defined as
\begin{equation} \label{curvature}
\kappa(u) = \frac{\left|\psi'(u) \zeta''(u)-\psi''(u) \zeta'(u)\right|}
{\left((\psi'(u))^2+( \zeta'(u))^2\right)^{3/2}}.
\end{equation}
The curvature vanishes at points where 
$\psi'(u) \zeta''(u)-\psi''(u) \zeta'(u)=0$.

%%e.g. Do Carmo p 25

\vs

\begin{defin}
A smooth device is said to be {\em locally nonlinear} if the curvature %of the characteristic
does not vanish identically
on any %positive-length %open (non-empty) subset/patch/
%portion of the curve/ % with positive length.
open %subset of the characteristic/
portion of its characteristic. %curve. 
%portion of the curve diffeomorphic to an interval.
\end{defin}

\vs

\noindent 
Here ``open'' is meant 
in the relative topology of the characteristic as a planar
1-manifold; in other words, the requirement is that the curvature does not vanish on any portion of the curve
diffeomorphic to an open interval.
%, that is, there is an interval where an affine relation of the form
%$\alpha\psi(u)+ \beta\zeta(u)=\gamma$ holds.
%[REMOVE?] It is very easy to check that, the other way round, 
%a
A device which is
not locally nonlinear has at least a portion of the characteristic which
is a line segment.

\vs

\begin{theor}
%Assume that 
If all devices of a smooth, uncoupled,
topologically nondegenerate circuit 
are locally nonlinear, then %. Then
the regular set ${\cal R}$ 
%the product ${\cal R}_c \times {\cal R}_l \times {\cal R}_r$ 
is open dense
in the homogeneous space $\h$.
%$\mathbb{H}_c \times \mathbb{H}_l \times \mathbb{H}_r$.
\end{theor}

\vs

\noindent {\bf Proof.} 
The fact that ${\cal R}$ is open
follows in a straightforward manner from Definition \ref{defin-reg}.
To show that it is also dense, it is enough to show that 
the sets ${\cal R}_c$, ${\cal R}_l$ and ${\cal R}_r$ are dense
in $\h_c$, $\h_l$ and $\h_r$, respectively. %, and the result follows.
%The non-trivial part of the claim concerns ${\cal R}_r$
%the latter 
%and is addressed below. Before that,
%note that 
Regarding ${\cal R}_c$ and ${\cal R}_l$, simply note that these are
the sets where all the components of
$\psi_c'(u_c)$ and $\zeta_l'(u_l)$ are non-zero. Assuming for instance
${\cal R}_c$ not to be dense in $\h_c$, there would
exist an open set in $\h_c$ where at
least one of the components of $\psi_c'(u_c)$, say $\psi_{c_i}'(u_{c_i})$,
should vanish. By taking a product of open intervals within 
that open set, not only $\psi_{c_i}'$ but also $\psi_{c_i}''$ 
would vanish on an open interval. In light of (\ref{curvature}),
this would imply
that the curvature of the characteristic of the $i$-th capacitor
vanishes on an interval,
 against the local nonlinearity
assumption. The same reasoning applies to show that ${\cal R}_l$ is 
dense in $\h_l$.

Assume now that ${\cal R}_r$ is not dense
in $\mathbb{H}_r$. This is equivalent to 
the assumption that the identity $K(u_r)=0$ (cf.\ %with $K$ defined
\eqref{kir-nonl})
holds on some open set within $\mathbb{H}_r$. 
Pick any resistive branch (say number 1, w.l.o.g.).
By restricting the aforementioned open set if necessary we may guarantee that either $p_{r_1}(u_{r_1})=\psi_{r_1}'(u_{r_1})$
or $q_{r_1}(u_{r_1})=\zeta'_{r_1}(u_{r_1})$ (we choose the latter, again w.l.o.g.\ as detailed later)
does not vanish on an interval $I_1$. 
%(within... product topology $I_1 \times I_{resto}$... explicar). 
The key fact is that the Kirchhoff polynomial $\tilde{K}(p,q)$ 
is homogeneous of degree one in $p_{r_1}$, $q_{r_1}$, and therefore 
we may divide by $q_{r_1}$ to get 
\ifthenelse{\boolean{ieee}}{\begin{eqnarray} 
%&& 
K_1(u_r)=\frac{K(u_r)}{q_{r_1}(u_{r_1})}= y_{r_1}(u_{r_1}) K_{11}(u_{r_2}, \ldots,
u_{r_{m_r}}) + %\hspace{10mm} 
\nonumber \\  %&& \mbox{} \hspace{30mm} 
+ K_{12}(u_{r_2}, \ldots,
u_{r_{m_r}}) \hspace{13mm}
\label{quot1}
\end{eqnarray}}{\begin{equation} \label{quot1}
K_1(u_r)=\frac{K(u_r)}{q_{r_1}(u_{r_1})}= y_{r_1}(u_{r_1}) K_{11}(u_{r_2}, \ldots,
u_{r_{m_r}}) + K_{12}(u_{r_2}, \ldots,
u_{r_{m_r}})
\end{equation}}
with $y_{r_1}=p_{r_1}/q_{r_1}$. Note that either $K_{11}$ or $K_{12}$ 
(but not both) might
be absent in the expression above 
%themselves vanish identically 
for topological reasons: 
e.g.\ if the first resistor is present in all
proper trees then all terms of $K$
include $p_{r_1}$ (and none $q_{r_1}$) as a factor, meaning that the
$K_{12}$ term would not be present; in the
dual case (namely, when all terms include $q_{r_1}$)
the identity (\ref{quot1}) would amount to $K_1=K_{12}$.
Including these two scenarios is necessary in order to guarantee
that there is no loss of generality in the non-vanishing assumption
on $q_{r_1}$ made above.

By construction and with the restriction mentioned above, 
the quotient in (\ref{quot1}) vanishes on the same
set as $K(u_r)$ and, therefore, we also have
%\begin{equation}\label{K1u1}
$\partial K_1/\partial u_{r_1} = 0$
%\end{equation}
on the same set. Now let us first assume 
that the $K_{11}$ term is indeed present
in (\ref{quot1}). From the vanishing of the first partial derivative
we get
\begin{equation}\label{K1u1bis}
y_{r_1}'(u_{r_1}) K_{11}(u_{r_2}, \ldots, u_{r_{m_r}})=0.
\end{equation}
Assuming the factor $y_{r_1}'(u_{r_1})$
to vanish on an open interval within the aforementioned $I_1$, we would get
%the identity 
$\psi_{r_1}'\zeta''_{r_1}-\psi_{r_1}''\zeta'_{r_1}=0$
there, against the local nonlinearity assumption on the first resistor.
It then follows from (\ref{K1u1bis}) that 
$K_{11}(u_{r_2}, \ldots, u_{r_{m_r}})$ must vanish identically on some
open set.
Should, on the other hand, the $K_{11}$ term be absent
from (\ref{quot1}), it would follow trivially that $K_1=K_{12}$ and 
%therefore 
the latter would vanish on the same (restricted) open set
where $K_1$ and $K$ do.

One way or another we get $K_{1i}(u_{r_2}, \ldots, u_{r_{m_r}})=0$ on some open
set,
either for $i=1$ or $i=2$. But again this is a multihomogeneous
polynomial on each pair of variables $p_j$, $q_j$ and the same reasoning
applies recursively. 
This way the argument can be repeated until some $y_{r_k}'$ vanishes
on some open subinterval, which contradicts the local nonlinearity assumption
on all resistors. 
This shows that ${\cal R}_r$ is indeed
dense in $\h_r$ and the proof is complete.
\hfill $\Box$

%Elaborate on the implications, impasse set... Linear/nonlinear distinction (THIS TO THE ABSTRACT).  Genericity of locally nonlinear maps.

%Remark: a single (locally) linear device may be enough to break the density property above, e.g. conductance
%in a proper tree with only one resistor (VdPol: the presence of other eventual (appropriately placed)
%nonlinear resistors would not interfere)

\subsection{On the manifold structure of the constraint set. Quasilinear reduction}
\label{subsec-manifold}

We finish this section with some brief remarks on the structure of the constraint
set ${\cal M}$ near impasse points. Let us emphasize the rather obvious 
fact that the non-singularity
of (\ref{matrix1}) is not a {\em necessary} condition
for the constraint set ${\cal M}$ defined by (\ref{core-alg}) 
to be a manifold. In greater generality, this set
would have a manifold structure near a given point
if the map in the left-hand side of 
this equation is (locally) a submersion, that is, if the matrix of partial derivatives
\begin{eqnarray} \label{derivGrande}
\begin{pmatrix} A_c^{\perp} A_l \psi_l'(u_l) & 0 & A_c^{\perp} A_r\psi_r'(u_r)  \\
0 & B_l^{\perp} B_c\zeta_c'(u_c) & B_l^{\perp} B_r\zeta_r'(u_r) 
\end{pmatrix}
\end{eqnarray}
has maximal rank $m_r$. Even if for brevity we state the following result without proof, it is
worth noting that the maximal rank condition on (\ref{derivGrande})
is met
% can be shown to hold 
in the setting
described %in Proposition \ref{propo-Mmanifold} 
below.

%Examine impasse points ((\ref{matrix1}) singular, or 
%$\psi_c'(u_c) =0$, or $\zeta_l'(u_l)=0$

\vs

\begin{propo} \label{propo-Mmanifold}
Assume that, at a given $(u_c, u_l, u_r) \in {\cal M}$,
all components of $\psi_l'(u_l)$ and $\zeta_c'(u_c)$ do not vanish, 
%for future use: these conditions mean that inductors are current-controlled 
%and capacitors voltage-controlled
and
that the matrix
\begin{eqnarray}  \label{maxrankr}
\begin{pmatrix} A_r\psi_r'(u_r) \\
B_r\zeta_r'(u_r)
\end{pmatrix}
\end{eqnarray}
has maximal rank $m_r$. Then ${\cal M}$ is locally a manifold near $(u_c, u_l, u_r)$.
\end{propo}

\vs
\begin{versionC}
\color{red}
\noindent {\bf Proof.} The first step of the proof proceeds as in Theorem \ref{th-reg}: 
premultiplying the matrix 
\begin{eqnarray} \label{matrixEnlarged}
%\cork 
\begin{pmatrix} A_c & A_l \psi_l'(u_l) & 0 & 0 & A_r\psi_r'(u_r) \\ 
0 & 0 & B_c\zeta_c'(u_c) & B_l & B_r\zeta_r'(u_r)\end{pmatrix}
% = \cork  \begin{pmatrix} A_c^{\perp} A_l \psi_l'(u_l)  & 0 & A_c^{\perp}A_r\psi_r'(u_r) \\ 
%0 & B_l^{\perp}B_c\zeta_c'(u_c) & B_l^{\perp}B_r\zeta_r'(u_r)
%\end{pmatrix}, \ \ 
\end{eqnarray}
by the block-diagonal one block-diag($\tilde{A}$, $\tilde{B}$)
from (\ref{ACBLbis}),
%by the nonsingular matrix (\ref{matrixAux})  we get the identity
we easily get
\ifthenelse{\boolean{ieee}}{
\begin{eqnarray*}
%&& \begin{pmatrix}A_c^{\tra} \vspace{0.5mm} & 0 \\ A_c^{\perp} & 0 \\ 
%0 & B_l^{\perp} \vspace{0.5mm} \\ 0 & B_l^{\tra} \end{pmatrix}
%\begin{pmatrix} A_c & A_l \psi_l'(u_l) & 0 & 0 & A_r\psi_r'(u_r) \\ 
%0 & 0 & B_c\zeta_c'(u_c) & B_l & B_r\zeta_r'(u_r)\end{pmatrix}
\hspace{-4mm} 
\begin{pmatrix} A_c^{\tra} A_c & * & * & * & * \\ %A_c^{\tra}A_r\psi_r'(u_r) \\ 
0 & A_c^{\perp} A_l \psi_l'(u_l) & 0 & 0 & A_c^{\perp}A_r\psi_r'(u_r) \\ 
0 & 0 & B_l^{\perp}B_c\zeta_c'(u_c) & 0 & B_l^{\perp}B_r\zeta_r'(u_r)\\
0 & * & * & B_l^{\tra} B_l & * %B_l^{\tra} B_r\zeta_r'(u_r)
\end{pmatrix}
\end{eqnarray*}}{
\begin{eqnarray*}
%&& \begin{pmatrix}A_c^{\tra} \vspace{0.5mm} & 0 \\ A_c^{\perp} & 0 \\ 
%0 & B_l^{\perp} \vspace{0.5mm} \\ 0 & B_l^{\tra} \end{pmatrix}
%\begin{pmatrix} A_c & A_l \psi_l'(u_l) & 0 & 0 & A_r\psi_r'(u_r) \\ 
%0 & 0 & B_c\zeta_c'(u_c) & B_l & B_r\zeta_r'(u_r)\end{pmatrix}
% =   \hspace{70mm} \\ && \hspace{45mm} =  
\begin{pmatrix} A_c^{\tra} A_c & * & * & * & * \\ %A_c^{\tra}A_r\psi_r'(u_r) \\ 
0 & A_c^{\perp} A_l \psi_l'(u_l) & 0 & 0 & A_c^{\perp}A_r\psi_r'(u_r) \\ 
0 & 0 & B_l^{\perp}B_c\zeta_c'(u_c) & 0 & B_l^{\perp}B_r\zeta_r'(u_r)\\
0 & * & * & B_l^{\tra} B_l & * %B_l^{\tra} B_r\zeta_r'(u_r)
\end{pmatrix},
\end{eqnarray*}}
where $*$ denotes entries which are not relevant to our
present purposes. Using the nonsingular
blocks $A_c^{\tra} A_c$ and $B_l^{\tra} B_l$ one can check that 
%the second factor in the left-hand side of the latter equation, namely,
and (\ref{derivGrande}) and (\ref{matrixEnlarged}) have the same corank.
Therefore, the maximal rank condition on (\ref{derivGrande}) can be equivalently
examined in terms of %the matrix in 
(\ref{matrixEnlarged}).

Now let $v=(v_1, v_2, v_3, v_4, v_5)$ be a vector in the kernel of the matrix (\ref{matrixEnlarged}) and
write, for notational simplicity in what follows, $P_l=\psi_l'(u_l)$, $Q_c=\zeta_c'(u_c)$,
$P_r =\psi_r'(u_r)$ and $Q_r=\zeta_r'(u_r)$. In light
of the orthogonality of the cycle and cut spaces (see e.g.\
\cite{bollobas}), the relations 
\begin{equation} \label{equat1}
v_1=B_c^{\tra}w_1, \ v_2 = P_l^{-1}B_l^{\tra}w_1, \ v_3 = Q_c^{-1}A_c^{\tra}w_2, \ v_4 = A_l^{\tra}w_2
\end{equation}
and
$P_r v_5 = B_r^{\tra}w_1, \
Q_r v_5 = A_r^{\tra}w_2$
must hold for certain vectors $w_1$, $w_2$. Note that, by construction, the matrix $T=P_r^2+Q_r^2$
is non-singular and then 
$v_5$ can be also explicitly written in terms of $w_1$ and $w_2$ as
\begin{equation} \label{equat2}
v_5=T^{-1}\left(P_rB_r^{\tra}w_1 + Q_r A_r^{\tra}w_2\right).
\end{equation}
Additionally, since $P_r$ and $Q_r$ are diagonal and hence commute, we have 
$Q_rP_r v_5=P_rQ_r v_5$
and then $Q_rB_r^{\tra}w_1=P_r  A_r^{\tra}w_2$.
This means that $(w_1, \ w_2) \in \ke (-Q_rB_r^{\tra} \ \ P_r  A_r^{\tra})$. The latter matrix has maximal rank by
hypothesis, so that its kernel has dimension $m-m_r = m_c + m_l$.

Finally, since the kernel of (\ref{matrixEnlarged}) 
can be written in terms of $w_1$ and $w_2$ via
(\ref{equat1}) and (\ref{equat2}), and as indicated above $w_1$, $w_2$ lie on a space
of dimension $m_c + m_l$,
the dimension of the kernel of (\ref{matrixEnlarged}) cannot be greater than $m_c + m_l$.
But the matrix  (\ref{matrixEnlarged})  has order $m \times (m + m_c + m_l)$, 
and therefore 
its rank may not be less than $m + m_c + m_l-(m_c+m_l)=m$; we then get that this rank indeed attains its
maximum possible value, $m$, as we aimed to show.
\hfill $\Box$

\vs

\color{black}
\end{versionC}

\noindent The maximal rank assumption on the matrix (\ref{maxrankr}) 
%is relevant in practice because it
can be shown to express the transversality of the projection
$(i_c, v_c, i_l, v_l, i_r, v_r) \to (i_r, v_r)$ (restricted
to the linear space defined by Kirchhoff laws) to the 
characteristic manifold ${\cal C}_r$. Find details in this
regard in \cite{smale}.
%And even if we omit a detailed discussion for the sake of brevity,
 Proposition  \ref{propo-Mmanifold} is useful from a dynamical perspective because
the manifold structure of ${\cal M}$ still allows
for a quasilinear description of the dynamics. Now this might not
be possible in terms of $u_c$, $u_l$ as in (\ref{state}), but 
%still 
it will be %possible 
in terms of some $m_c+m_l$ homogeneous variables from within the
vector $(u_c, u_l, u_r)$. Just for illustrative purposes,
an elementary example can be given in terms of 
%quasilinear reduction holding generically, because max rank: some $u_r$'s, some $u_c$'s etc. Bring equations: 
%\begin{subequations} \label{core-diffRepe}
%\begin{eqnarray}
%\psi_c'(u_c)u_c' & = & -A_c^-( A_l \psi_l(u_l) + A_r\psi_r(u_r))\\
%\zeta_l'(u_l)u_l' & = & -B_l^- (B_c\zeta_c(u_c) + B_r\zeta_r(u_r))
%\end{eqnarray}
%\end{subequations}
%and
%\begin{subequations} \label{core-algRepe}
%\begin{eqnarray}
%A_c^{\perp} \left( A_l \psi_l(u_l) + A_r\psi_r(u_r)\right) & = & 0 \\
%B_l^{\perp} \left( B_c\zeta_c(u_c) + B_r\zeta_r(u_r)\right) & = & 0.
%\end{eqnarray}
%\end{subequations}
%From the latter some $u_l$'s and $u_r$'s in terms of the others (and the same for $c$'s: CAREFUL, ALL ``AT THE SAME TIME'')
%say ${u_l}_2 = f_1 ({u_l}_1, {u_r}_1, {u_c}_1)$, ${u_r}_2=f_2({u_l}_1, {u_r}_1,% {u_c}_1)$, 
%${u_c}_2=f_3({u_l}_1, {u_r}_1, {u_c}_1)$,
%
%equations for ${u_l}_1$ and ${u_c}_1$ clear, but what about ${u_r}_1$? MAYBE JUST INSERT f1 f2 and f3 in
%the equations above and get a quasilinear system in terms of the variables with subscript 1. Illustrate with an example (VdPol). MIGHT BE ENOUGH
%
%In the example insert 
%$v_c=\zeta_r(u_r)$ 
%OR IN GREATER GENERALITY TO SHOW ALL IMPASSE POINTS IN THE REDUCTION
(\ref{vdp1}): even near an impasse point defined by the condition
$\zeta_r'(u_r)=0$, the constraint set (given by $v_c=\zeta_r(u_r)$)
is a manifold where a quasilinear reduction is still feasible,
now in terms of $u_l$, $u_r$. Note that impasse points are captured in the leading coefficients
of the reduction, which has the form
\begin{subequations}\label{vdp1ql}
\begin{eqnarray} 
%p(u_c)
C\zeta_r'(u_r)u_r' & = & \psi_l(u_l) - \psi_r(u_r)\\
%q(u_l) 
\zeta_l'(u_l)u_l' & = & -\zeta_r(u_r).
\end{eqnarray}
\end{subequations}

\section{Memristors}
\label{sec-mem}

In this section we briefly 
show how to extend the previous approach to circuits with 
memristors, a family of devices
which has attracted a lot of attention in Electronics 
in the last decade, following the results reported %in 2008 
in the paper
\cite{strukov}.
By means of a specific example we show the form that the models take and,
in particular, how
the homogeneous formalism makes it possible to frame in the same context
two problems considered in \cite{corinto2016, corinto2017}.
%; this analysis suggests the presence
%of a subtle form of duality in memristor circuits, which is probably worth a more detailed
%analysis in the future.
%For the sake of brevity we only sketch the results.

\subsection{Homogeneous modelling of circuits with memristors}
 
A memristor is any electronic device characterized by a nonlinear relation between
the charge $\sigma$ and the magnetic flux $\varphi$.
Under the assumption that this relation is smooth,
%(more precisely, that the characteristic is a smooth planar curve)
we may proceed as in Section \ref{sec-homogmodels} to describe 
%a smooth memristive 
this characteristic in terms of a homogeneous variable $u$ in the form
%\begin{subequations}
\begin{eqnarray}
    \sigma=\psi(u), \ \varphi= \zeta(u). \label{memristor}
\end{eqnarray}
%\end{subequations}
Under the obvious nonvanishing assumptions, either the 
{\em memristance} $\zeta'(u)/\psi'(u)$
or the {\em memductance} $\psi'(u)/\zeta'(u)$ 
are well-defined at any $u$. In greater generality,
the {\em homogeneous memristance} reads as $(\psi'(u): \zeta'(u))$.

With the addition of memristors, the homogeneous model (\ref{core}) takes the form
\ifthenelse{\boolean{ieee}}{\begin{subequations}
\begin{eqnarray}
\hspace{-7.5mm}A_m \psi_m'(u_m) u_m' \hspace{-0.5mm}+\hspace{-0.5mm} A_c \psi_c'(u_c) u_c' \hspace{-0.5mm}+\hspace{-0.5mm} A_l \psi_l(u_l) \hspace{-0.5mm}+\hspace{-0.5mm} A_r\psi_r(u_r) & \hspace{-3.2mm}=\hspace{-3.2mm} & 0 \\
\hspace{-10mm}B_m\zeta_m'(u_m)u_m' \hspace{-0.5mm}+\hspace{-0.5mm} B_c\zeta_c(u_c) \hspace{-0.5mm}+\hspace{-0.5mm} B_l\zeta_l'(u_l)u_l' \hspace{-0.5mm}+\hspace{-0.5mm} B_r\zeta_r(u_r) & \hspace{-3.2mm}=\hspace{-3.2mm} & 0,\hspace{3.9mm}
\end{eqnarray}
\end{subequations}}{\begin{subequations}
\begin{eqnarray}
A_m \psi_m'(u_m) u_m' + A_c \psi_c'(u_c) u_c' + A_l \psi_l(u_l) + A_r\psi_r(u_r) & = & 0 \ \\
B_m\zeta_m'(u_m)u_m' + B_c\zeta_c(u_c) + B_l\zeta_l'(u_l)u_l' + B_r\zeta_r(u_r) & = & 0,
\end{eqnarray}
\end{subequations}}
with the vector-valued maps $\psi_m$ and $\zeta_m$ joining together the contributions
of the different memristors. We illustrate below the form that these equations
may take in practice.

\subsection{Example}
\label{subsec-exmem}

The memristor-capacitor circuit displayed in Fig.\ 4 
%Figure \ref{fig-MC} 
is analyzed, under different assumptions,
in \cite{corinto2016, corinto2017}. We show below how 
our approach
%the approach introduced in this paper 
makes it possible
to accommodate both analyses in a single, unifying framework, unveiling in addition
some symmetry properties
which underly this example and possibly other memristive circuits.  We assume for simplicity 
that %in all cases 
the capacitor is linear, %one 
with $C=1$.

\ifthenelse{\boolean{ieee}}{
\begin{figure}[h]
%\vspace{2mm}
\centering
%\mbox{} \hspace{-6mm} 
%\parbox{2.0in}{
%\vspace{5.5mm}
\epsfig{figure=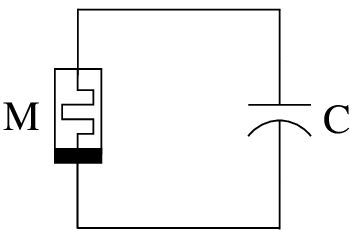, width=0.17\textwidth}%, angle=270}
%}
%\ \mbox{} \hspace{-25mm} \
%\vspace{1mm}
\caption{Memristor-capacitor circuit.}
\label{fig-MC}
%\vspace{1mm}
\end{figure}}{
\begin{figure}[h]
%\vspace{2mm}
\centering
%\mbox{} \hspace{-6mm} 
%\parbox{2.0in}{
%\vspace{5.5mm}
\epsfig{figure=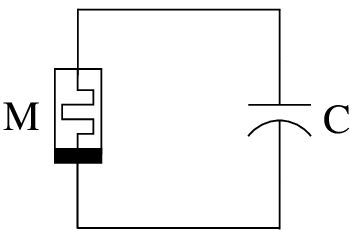, width=0.24\textwidth}%, angle=270}
%}
%\ \mbox{} \hspace{-25mm} \
%\vspace{1mm}
\caption{Memristor-capacitor circuit.}
\label{fig-MC}
%\vspace{1mm}
\end{figure}}

%\noindent 
In \cite{corinto2016} %flux-charge
the memristor is assumed to be flux-controlled, with a cubic characteristic which can be written
in the form $\sigma_m = -\varphi_m + \varphi_m^3$. %(we fix for simplicity the parameter values 
%$a=b=1$ in the characteristic $-a\varphi_m + b\varphi_m^3$ actually considered there). 
Two
stability changes are reported in that paper to occur along a line of equilibria and 
for the flux values $\varphi_m = \pm \sqrt{1/3}$; more precisely,
this circuit can be shown to undergo
two transcritical bifurcations without parameters by checking that it satisfies
the general requirements characterizing this bifurcation
in \cite{tbwp}. %(see also \cite{fiedler00a, liebscherbook} in this regard). 
By contrast, in \cite{corinto2017} %bif without parameters
the memristor is assumed to have the dual charge-controlled
form  $\varphi_m=-\sigma_m + \sigma_m^3$, which is responsible for the presence of 
two impasse manifolds, defined by the charge values
$\sigma_m = \pm \sqrt{1/3}$,
 where trajectories collapse in finite time with infinite speed.
%This is a well-known phenomenon in nonlinear
%circuit theory, which can be traced back to the seminal works Sastry and Desoer... Chua
%and Deng.

What we want to examine is the reason for the dual characteristics above
to yield these two qualitative phenomena. Note that in the framework of 
\cite{corinto2016, corinto2017} two different models must be used, because
of the different control variables involved in the memristor;
indeed, in the former case the circuit equations are formulated in \cite{corinto2016}
in terms of the %memristor 
flux, and necessarily in terms of the charge in
%whereas in the latter case the circuit model
%is necessarily formulated in terms of the charge (cf.\ 
\cite{corinto2017}.
Instead, a single reduction applying to both contexts can be obtained from the
homogeneous framework, making it possible to formulate a single model in terms
of one and the same homogeneous variable $u_m$ for the memristor 
(for the capacitor, because of its linear %(and non-degenerate) 
nature,
we may choose $v_c$, $\sigma_c$ or even 
a homogeneous variable $u_c$).
%(by contrast,
%the linear nature of the capacitor allows for the use of any capacitor variable
%in the model; we choose the voltage $v_c$ but also the charge $\sigma_c$ or even
%a homogeneous variable $u_c$ would also be admissible). 

Specifically, the equations
for the circuit in Fig.\ 4 %ure \ref{fig-MC} 
can be written, using
an homogeneous description of the memristor
(cf.\ (\ref{memristor})), 
as
\begin{subequations} \label{mcex}
\begin{eqnarray}
%    \sigma_m=\psi
p(u_m)u_m' - v_c' & = & 0\\
q(u_m)u_m' & = & -v_c,
\end{eqnarray}
\end{subequations}
with $p(u_m)=\psi_m'(u_m)$, $q(u_m)=\zeta_m'(u_m)$. Here we need no assumption on controlling
variables in the memristor. In particular, 
denoting $\chi(u_m)=-u_m + u_m^3$,
the two cases considered in \cite{corinto2016, corinto2017} are accommodated in this model
just by setting $\psi_m = \chi$ and $\zeta_m=\mathrm{id}$ 
%%\begin{subequations}
%%\begin{eqnarray}
%    $\sigma_m  =  \psi_m(u_m) = -u_m + u_m^3$, $\varphi_m  =  \zeta_m(u_m) = u_m$,
%%\end{eqnarray}
%%\end{subequations}
(with $p(u_m)=\chi'(u_m)=-1+3u_m^2$, $q(u_m)=1$)
to model the flux-controlled context of \cite{corinto2016},
and
$\psi_m=\mathrm{id}$, $\zeta_m=\chi$ 
%%\begin{subequations}
%%\begin{eqnarray}
%    $\sigma_m  =  \psi_m(u_m) = u_m$, $\varphi_m =  \zeta_m(u_m) = -u_m + u_m^3$,
(yielding $p(u_m)=1$, $q(u_m)=\chi'(u_m)$) %-1+3u_m^2$)
%%\end{eqnarray}
%%\end{subequations}
for the charge-controlled setting of \cite{corinto2017}. 
%%(where  $\varphi_m=-q_m + q_m^3$). 
%%In the former case we have $p(u_m)=-1+3u_m^2$, $q(u_m)=1$ and,
%%in the latter, $p(u_m)=1$, $q(u_m)=-1+3u_m^2$. 

Regardless of the actual form of the memristor characteristic, it is clear
from (\ref{mcex}) that this system has a line of equilibria 
defined by $v_c=0$. %Denoting $p(u_m)=\psi_m'(u_m)$, $q(u_m)=\zeta_m'(u_m)$,
The linearization of (\ref{mcex}) at any equilibrium point
is defined by the matrix pencil
\begin{eqnarray}
\lambda \begin{pmatrix} p(u_m) & -1 \\ q(u_m) & 0
\end{pmatrix} +
\begin{pmatrix} 0 & 0 \\
0 & 1
\end{pmatrix},
\end{eqnarray}
whose eigenvalues are given by the roots of the polynomial $\lambda(\lambda q(u_m) + p(u_m))$; these 
are $\lambda=0$ and 
%\begin{equation}
$\lambda = -p(u_m)/q(u_m).$ %\label{eigen} 
%\end{equation}
Worth remarking is the fact that the null eigenvalue reflects that equilibrium
points are not isolated but define a line, a phenomenon which is well-known to happen
%systematically 
in the presence of a memristor (see \cite{tbwp} and references therein).

Now, the zeros of $p$ and of $q$ in each one
of the cases defined by the characteristics
of \cite{corinto2016, corinto2017}
are located
at $u_m = \pm \sqrt{1/3}$. The zeros of $p$ in the first setting 
define a second null eigenvalue in the
%matrix 
pencil spectrum, which is responsible for the transcritical bifurcation
without parameters; in turn,
the zeros of $q$ in the second 
case yield an infinite eigenvalue in the pencil, which results in the aforementioned
impasse phenomenon. 
The key remark is that the homogeneous model (\ref{mcex}) %is able to
accommodates simultaneously both contexts, capturing the intrinsic
symmetry of both problems; actually, this framework
%expression depicted in (\ref{eigen}) 
(specifically, the expression for the second eigenvalue)
makes it apparent %, in the light of (\ref{eigen}),
that the nontrivial eigenvalue is transformed by the 
relation $\lambda \to %\leftrightarrow 
1/\lambda$ when 
the expressions defining $p$ and $q$ are interchanged. Now it becomes clear that stability changes
in the first setting, due to
the transition of an eigenvalue through zero in the transcritical bifurcation without parameters, 
correspond in the second context to a sign change in the eigenvalue 
owing to its divergence through $\pm \infty$.
%This spectral inversion and, more generally, the relation between both qualitative
%phenomena, seems worth being
%analyzed in more detail, a task which is left for future study.

\section{Concluding remarks}
\label{sec-con}

We have extended in this paper the homogeneous approach of \cite{homoglin} to 
uncoupled nonlinear %electrical 
circuits, possibly including memristors, under a smoothness
assumption on all devices.
%Using the formalism
%of sheaf theory, we have provided a construction 
%of independent mathematical interest
%which identifies smooth planar curves (describing the characteristics of individual
%devices in the circuit context) with classes of equivalent submersions, much as in the linear setting
%the projective construction of \cite{homoglin} identifies linear devices with equivalent
%linear forms. 
This framework leads to a 
new circuit model, displayed in (\ref{core00}) (find details 
in subsection \ref{subsec-main}, cf.\ (\ref{core})), which, involving only one state variable per branch, 
retains the %full 
generality of larger size
model families such as those arising in the tableau approach.
From the modelling perspective, worth emphasizing is the fact that
the homogeneous
model (\ref{core00}) particularizes to classical models in restricted scenarios 
in which some devices %are known to 
admit global descriptions in terms of the current, voltage, charge or flux; 
these %classical 
contexts are captured %simply obtained 
by appropriate choices of the
maps $\psi_r$, $\zeta_r$, $\psi_c$, etc.\ in (\ref{core00}). %This way we avoid 
Broadly, the homogeneous approach avoids the need to assume the %global 
existence of
such classical %current/voltage/charge/flux-controlled 
descriptions, which entail a loss of generality
in the formulation and the 
reduction of circuit models.
We have also briefly indicated how to extend the approach in order 
to accommodate controlled sources
and coupling effects.

Our results make it possible to address in detail 
certain analytical problems such as the state-space problem: 
%for nonlinear circuits:
%in nonlinear circuit theory: 
in this direction, we have provided
a full circuit-theoretic
characterization of the so-called regular manifold of topologically nondegenerate (index one)
circuits,
holding without any restriction on controlling variables of individual devices. We have also
proved that for so-called 
locally nonlinear problems the regular set is open dense in the homogeneous space, capturing
a subtle qualitative distinction between nonlinear (in the strict sense) and linear circuits,
since for the latter all points are known to be simultaneously 
regular or singular (the latter yielding higher index models in well-posed cases).
The homogeneous approach %may be expected to 
should be of help in other
analytical problems in circuit theory in the future.
%The extension of this approach to circuits with full coupling effects and multiterminal devices, 
%as well as to distributed circuits, is in the scope of future research.

%\vspace{-3mm}

\ifthenelse{\boolean{ieee}}{}{\end{document}}

\end{document}